\documentclass[twocolumn,superscriptaddress,amsmath,apswey,nofootinbib]{revtex4-2}

\usepackage{hyperref}
\usepackage{graphicx}
\usepackage{amsmath,amsfonts}
\usepackage{dcolumn}
\usepackage{bm}
\usepackage{color}
\usepackage{multirow}
\usepackage{physics}
\usepackage{subfigure}
\usepackage[utf8]{inputenc}
\usepackage{scrextend}
\usepackage{float}
\usepackage{bibentry}

\begin{document}

\title{Anisotropic, multiband, and strong-coupling superconductivity of the Pb$_{0.64}$Bi$_{0.36}$ alloy}

\author{Sylwia Gutowska}
\email{sylwia.gutowska@univie.ac.at}
\affiliation{University of Vienna, Faculty of Physics and Center for Computational Materials Science, Kolingasse 14-16, A-1090, Vienna, Austria}
\affiliation{AGH University of Krakow, Faculty of Physics and Applied Computer Science,
Aleja Mickiewicza 30, 30-059 Krakow, Poland}
\author{Karolina Górnicka}
\affiliation{Materials Science and Technology Division, Oak Ridge National Laboratory, Oak Ridge, Tennessee, United States}
\affiliation{Faculty of Applied Physics and Mathematics and Advanced Materials Centre, Gdansk University of Technology,
ul. Narutowicza 11/12, 80-233 Gdańsk, Poland}
\author{Pawe\l{} W\'ojcik}
\affiliation{AGH University of Krakow, Faculty of Physics and Applied Computer Science,
Aleja Mickiewicza 30, 30-059 Krakow, Poland}
\author{Tomasz Klimczuk}
\affiliation{Faculty of Applied Physics and Mathematics and Advanced Materials Centre, Gdansk University of Technology,
ul. Narutowicza 11/12, 80-233 Gdańsk, Poland}
\author{Bartlomiej Wiendlocha}
\email{wiendlocha@fis.agh.edu.pl}
\affiliation{AGH University of Krakow, Faculty of Physics and Applied Computer Science,
Aleja Mickiewicza 30, 30-059 Krakow, Poland}


\date{\today}

\begin{abstract}
This paper presents theoretical and experimental studies on the superconductivity of Pb${_{0.64}}$Bi$_{0.36}$ alloy, which is a prototype of strongly coupled superconductors and exhibits one of the strongest coupling under ambient pressure among the materials studied so far. The critical temperature, the specific heat in the superconducting state, and the magnetic critical fields are experimentally determined. Deviations from the single-gap s-wave BCS-like behavior are observed.
The electronic structure, phonons and electron-phonon interactions are analyzed in relation to the metallic Pb, explaining why the Pb-Bi alloy exhibits such a large value of the electron-phonon coupling parameter $\lambda \simeq 2$. 
Superconductivity is studied using the isotropic Eliashberg formalism as well as the anisotropic density functional theory for superconductors. We find that while Pb is a two-gap superconductor with well-defined separate superconducting gaps, in the Pb-Bi alloy an overlapped three-gap-like structure is formed with a strong anisotropy. 
Furthermore, the chemical disorder, inherent to this alloy, leads to strong electron scattering, which is found to reduce the critical temperature.
\end{abstract}


\maketitle

\section{Introduction}
Pb$_{1-x}$Bi$_x$ alloys are very strongly coupled classical superconductors that consist of the heaviest stable elements from the periodic table. 
The structural and superconducting properties of these materials have been the subject of extensive research, with the first reports dating back to the 1930s \cite{pbbi-history1}, and research continuing between the 1960s and 1980s~\cite{pbbi-history2,pbbi-history3,pbbi-history4,pbbi-history5,pbbi-history6}.
The phase diagram was found to reveal two distinct phases, $fcc$ and hexagonal (nearly $hcp$, so-called $\epsilon$-phase), and the latter forms above $x\simeq 20\%$ \cite{pbbi-history3,pbbi-history4}. 
The highest critical temperature observed was above 8\,K, found in the hexagonal phase when $x\simeq 30 - 40\%$ \cite{pbbi-history2}. 
Scanning tunneling microscopy suggested that the superconducting gap for $x=30\%$ is of BCS type \cite{pbbi-history6}. Magnetization measurements showed that magnetic critical fields increased with $x$, up to $H_{c1}=141$\,Oe and $H_{c2}=17.7$\,kOe for $x=40\%$ \cite{pbbi-history4}. Electron tunneling spectroscopy established that the critical temperature, superconducting gap, and the electron-phonon coupling (EPC) constant increased with $x$ reaching 8.9\,K, 1.84 meV and 2.13, respectively, for $x=0.35$. This was accompanied by a decrease in phonon frequencies \cite{pbbi-history5}. 
Recently, studies carried out in the full range of $x$ from 0 to 100\% with a concentration step of 5\% confirmed that the highest critical temperatures are obtained around $x\simeq 30\% - 40\%$  \cite{pbbi-history7}. 
The superconducting parameters were also determined for Pb-Bi microcubes \cite{pbbi-history8}, films \cite{pbbi-history17} and amorphous phase \cite{chen1971electron}.
 
Pb-Bi alloys because of their relatively high critical temperatures and magnetic fields are suitable for numerous practical applications. For example, they are widely used as superconducting solder and joints \cite{pbbi-history10}, capable of transmitting 1000\,A in a 1\,T field at 4.2 K \cite{pbbi-history11}. These alloys were also proposed to be used in superconducting phonon spectrometer \cite{pbbi-history12}. Recently, they have been extensively studied as possible nuclear coolants \cite{pbbi-history14,pbbi-history18} in generation IV nuclear reactors.

Despite the whole described experimental effort, there is a lack of theoretical attempts to understand why these alloys become such strong coupled superconductors and what the details are of the superconducting phase, including the structure of the superconducting gap. Heat capacity and critical field were analyzed by Daams {\it et al.} in 1979 on the basis of the isotropic Eliashberg function \cite{pbbi-history9} determined by tunneling spectroscopy. 
In 2012 De la Pe\~na-Seaman {\it et al.} performed calculations based on the isotropic Eliashberg formalism for the Pb-Bi and Pb-Tl alloys in the cubic phase for $x\leq 0.2$ \cite{pbbi-history19}, which was one of the first theoretical works in which spin-orbit coupling was included in the Eliashberg function calculations. 
Recently,  Xie {\it et al.} calculated the Eliashberg function for Pb-Bi films \cite{pbbi-history17}.

In this work, the superconductivity in Pb$_{0.64}$Bi$_{0.36}$ with $T_c$ = 8.6\,K and $\lambda \simeq 2$ is studied experimentally and theoretically. 
Magnetic susceptibility, resistivity, and specific heat are measured for polycrystalline samples to determine the parameters of the superconducting state. Clear indications of the anisotropic and strongly coupled nature of the superconducting state are found.  
Electronic and phonon structures are calculated using the Density Functional Theory (DFT). 
The electron-phonon interaction is analyzed focusing on the questions of how the addition of Bi enhances the electron-phonon coupling in the Pb-Bi alloy, making it a better superconductor than Pb, where $T_c$ = 7.2\,K and $\lambda \simeq 1.5$, and what the role of the spin-orbit coupling is.
The anisotropy of the EPC constant and superconducting gap is studied. 
Theoretical results are compared with experiments, giving a consistent picture of a multigap anisotropic superconductivity with one of the highest $\lambda$ and $2\Delta/k_BT_c$ recorded so far in bulk materials under ambient conditions.

\section{Materials and Methods}

The Pb$_{0.64}$Bi$_{0.36}$ polycrystalline sample was prepared by melting the required high-purity elements, i.e., Pb chunks and Bi pieces. We selected the Pb$_{0.64}$Bi$_{0.36}$ composition based on available phase diagrams and previous reports on the critical temperature and the Meissner fraction for Pb$_{1-x}$Bi$_{x}$. The elements in the 0.64: 0.36 atomic ratio were sealed inside a silica tube under a partial pressure of Ar. The ampule was heated to 350\textdegree C at a rate of 100\textdegree C/h, held at that temperature for 2 h, and quenched in water to room temperature. To ensure optimal mixing of the constituents, the sample nugget was annealed at 120\textdegree C for 10 days. The resulting material was silver in color and rather soft and malleable. 
The crystal structure and phase purity of the obtained sample were determined by LeBail refinement of room-temperature X-ray diffraction (XRD) data collected on a Bruker D2 Phaser diffractometer with Cu K$\alpha$ radiation and a LynxEye-XE detector. Due to the ductility of Pb$_{0.64}$Bi$_{0.36}$, for qualitative characterization, the sample had to be transformed into a plate. The small piece of material was converted to a plate form using a roller. Mechanical handling did not cause any sample contamination. Refinement of the XRD pattern was performed using {\sc diffrac.suite topas}.  
All physical property measurements were carried out using a Quantum Design Evercool II Physical Property Measurement System (PPMS). Magnetic data were collected in the temperature range 1.7–10 K under various applied magnetic fields. Specific heat measurements were performed between 1.95 and 300 K in zero field and under a magnetic field of 5 T, using the two-$\tau$ time relaxation method. The sample was attached to the $\alpha$-Al$_2$O$_3$ measurement platform by Apiezon N grease to ensure good thermal contact. The electrical resistivity was determined using a standard four-probe method, with four platinum wire leads spark-welded to the surface of the bar-shaped polished sample. Measurements were performed in the temperature range 1.9-300 K in magnetic fields up to 2.2~T.

Theoretical calculations of the electronic structure, phonons, and electron-phonon interaction functions were performed for Pb and Pb$_{0.64}$Bi$_{0.36}$ using the plane-wave pseudopotential method, implemented in the {\sc Quantum espresso} (QE) package~\cite{qe,qe2}. 
Elemental {\it fcc} Pb is discussed as a reference material and to reveal the mechanism of enhancement of $\lambda$ and $T_c$ after alloying with Bi.
To simulate the effect of alloying, the virtual crystal approximation (VCA) was used to generate the "average" mixed pseudopotential, based on ultra-soft pseudopotentials of Pb and Bi~\cite{pseudo}. Exchange-correlation effects were treated within the Perdew-Burke-Ernzerhof generalized gradient approximation (GGA) scheme~\cite{pbe}. 
To investigate the effect of the spin-orbit coupling (SOC), both the scalar-relativistic and fully-relativistic calculations were performed. The electronic structure was calculated on a grid of $16^{3}$ {\bf k}-points for the self-consistent cycle and $48^{3}$ for density of states (DOS) and Fermi surface (FS) calculations. 

{Since the electron-phonon properties of the investigated alloy firstly depend on its electronic structure, it was necessary to ensure that the VCA method correctly describes the electronic structure of the Pb-Bi alloy.}
The precision of the VCA approach in describing the electronic structure of Pb$_{0.64}$Bi$_{0.36}$ was verified using the Korringa-Kohn-Rostoker (KKR) method with the coherent potential approximation (CPA)~\cite{Ebert2011}, as implemented in the Munich {\sc SPR-KKR} band structure package~\cite{sprkkr}. This method is designed to describe the electronic properties of random alloys, and also allows one to study the disorder-induced electron scattering effects via analysis of the Bloch spectral density function (BSF) and residual resistivity.
Full-potential relativistic KKR-CPA calculations were performed on fine meshes of about $3\times 10^3$ and $2\times 10^5$ \textbf{k}-points for the self-consistent cycle and BSF calculations, respectively (the number of points is given in the irreducible part of the Brillouin zone). The angular momentum cutoff was set to $l_{\rm max} = 3$ and the Fermi level was determined using the Lloyd formula~\cite{Ebert2011}.
{Additionally, to investigate the effects of disorder-induced scattering, electronic lifetimes were deduced from the KKR-CPA spectral functions, and the residual electrical resistivity was calculated.
That was done using the Kubo-Greenwood formalism~\cite{kubo57,greenwood58,butler-kubo} and including the vertex corrections~\cite{Ebert2011,Ebert-pss}.}

As the agreement between KKR-CPA and VCA appeared to be very good {and due to a small difference in masses of Pb and Bi (the mass-disorder effects are expected to be negligible)} we proceeded with the phonon calculations in QE using density functional perturbation theory (DFPT) \cite{dfpt}, with atoms represented by the VCA pseudopotentials and having the weighted average atomic mass. 
The grid of 8$^{3}$ {\bf q}-points was used, resulting in the number of 50 independent dynamical matrices to be calculated. 
The electron-phonon interaction function was calculated based on the obtained phonon structure with the integrals on the Fermi surface calculated with the double delta smearing technique with the smearing parameter $\sigma=0.02$ Ry, and based on the electronic structure calculated on the mesh of $32^3$ {\bf k}-points.
Analogous calculations have been performed for $fcc$ Pb. We verified that to reproduce the experimental phonon spectrum of Pb, meshes of $16^3$ {\bf k}-points and $8^3$ {\bf q}-points are required.

The Eliashberg functions obtained were used to describe the thermodynamic properties of the superconducting state using the isotropic Eliashberg formalism \cite{Eliashberg1960} in the implementation described in Refs.~\cite{kuderowicz2022,kuderowicz2021}. 
Here, the semi-empirical retarded Coulomb pseudopotential parameter $\mu^*$ is used to describe the depairing effects.

Next, to obtain the fully {\it ab-initio} description of superconductivity in the materials studied, without the need to assume the value of $\mu^*$, the density functional theory for superconductors (SCDFT) in the decoupling approximation \cite{Oliveira1988,Luders2005,Kawamura2017,sctk}, as implemented in the {\sc Superconducting Toolkit} (SCTK) \cite{sctk} was used. 
In these calculations, the anisotropic gap equation is solved, and the screened Coulomb interaction parameter is calculated on the basis of the actual electronic structure, within the random phase approximation \cite{Kawamura2017,sctk}. 
For SCTK calculations, a shifted mesh of {\bf q}-points and norm-conserving pseudopotentials~\cite{pseudo2} were required, therefore we repeated the phonon computations on a shifted mesh of $8^3$ {\bf q}-points (60 and 80 non-equivalent dynamical matrices for Pb and Pb-Bi alloy, respectively).

\section{Experimental studies}
\begin{figure}[b]
\includegraphics[width=.47\textwidth]{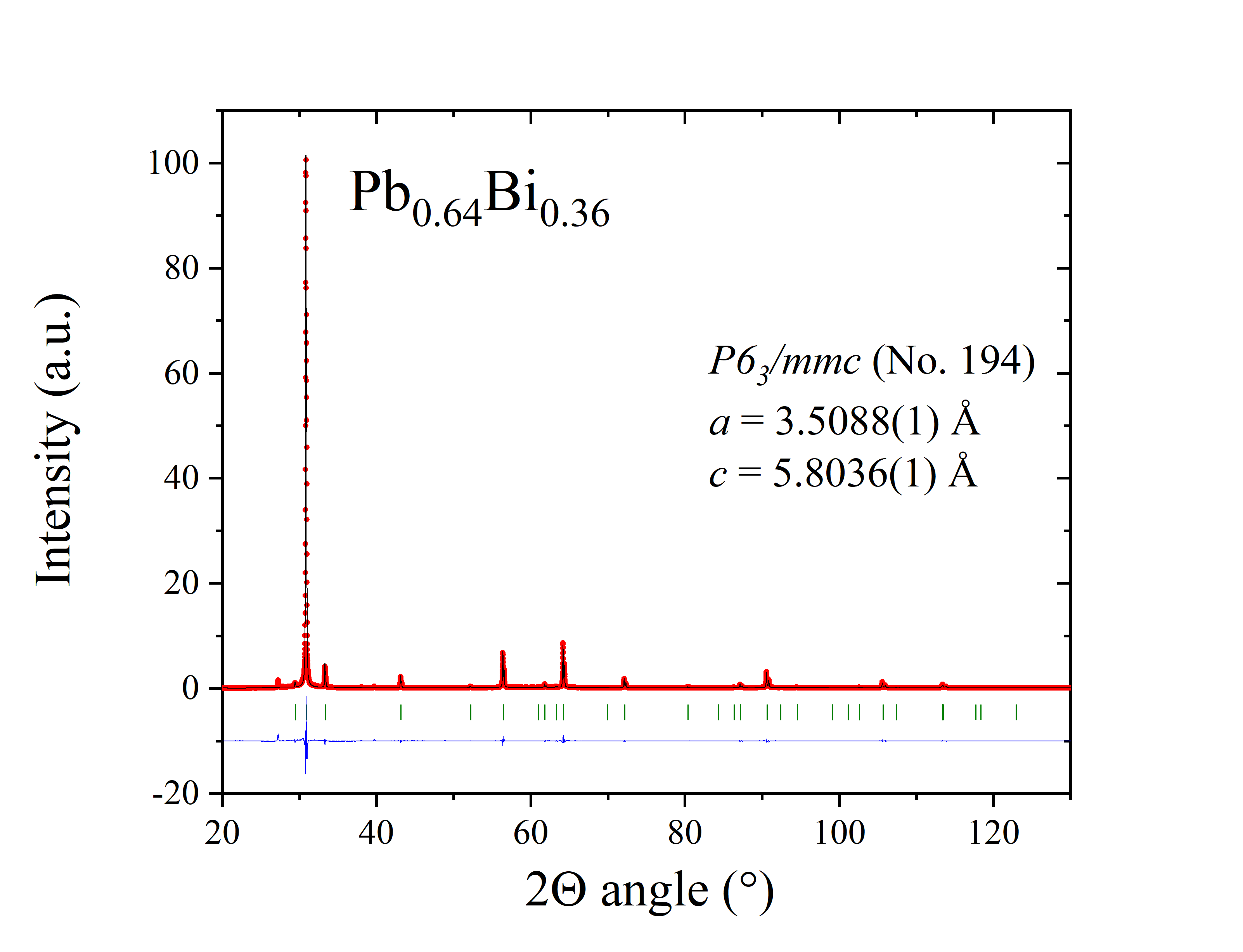}
\caption{The XRD pattern of Pb$_{0.64}$Bi$_{0.36}$ alloy (red points) together with the LeBail refinement profile (black solid line). The green vertical bars indicate the expected Bragg peak positions.}\label{fig:exp-xrd}
\end{figure}

The X-ray diffraction (XRD) pattern of Pb$_{0.64}$Bi$_{0.36}$ together with the LeBail refinement profile and Bragg positions are shown in Fig. \ref{fig:exp-xrd}. XRD pattern indicated good quality of the examined sample. In a more detailed analysis of the data, the P6$_3$/{\it mmc} phase (s.g. No.194) was refined with the LeBail method. The LeBail refinement to the diffraction pattern gave the lattice parameters $a = 3.5088(1)$ \AA~ and $c = 5.80356(1)$ \AA. The obtained values are in very good agreement with the previously published data for Pb$_{0.70}$Bi$_{0.30}$\cite{Rasmussen_Lundtoft_1987}. 

 \begin{figure}[b]
\includegraphics[width=.47\textwidth]{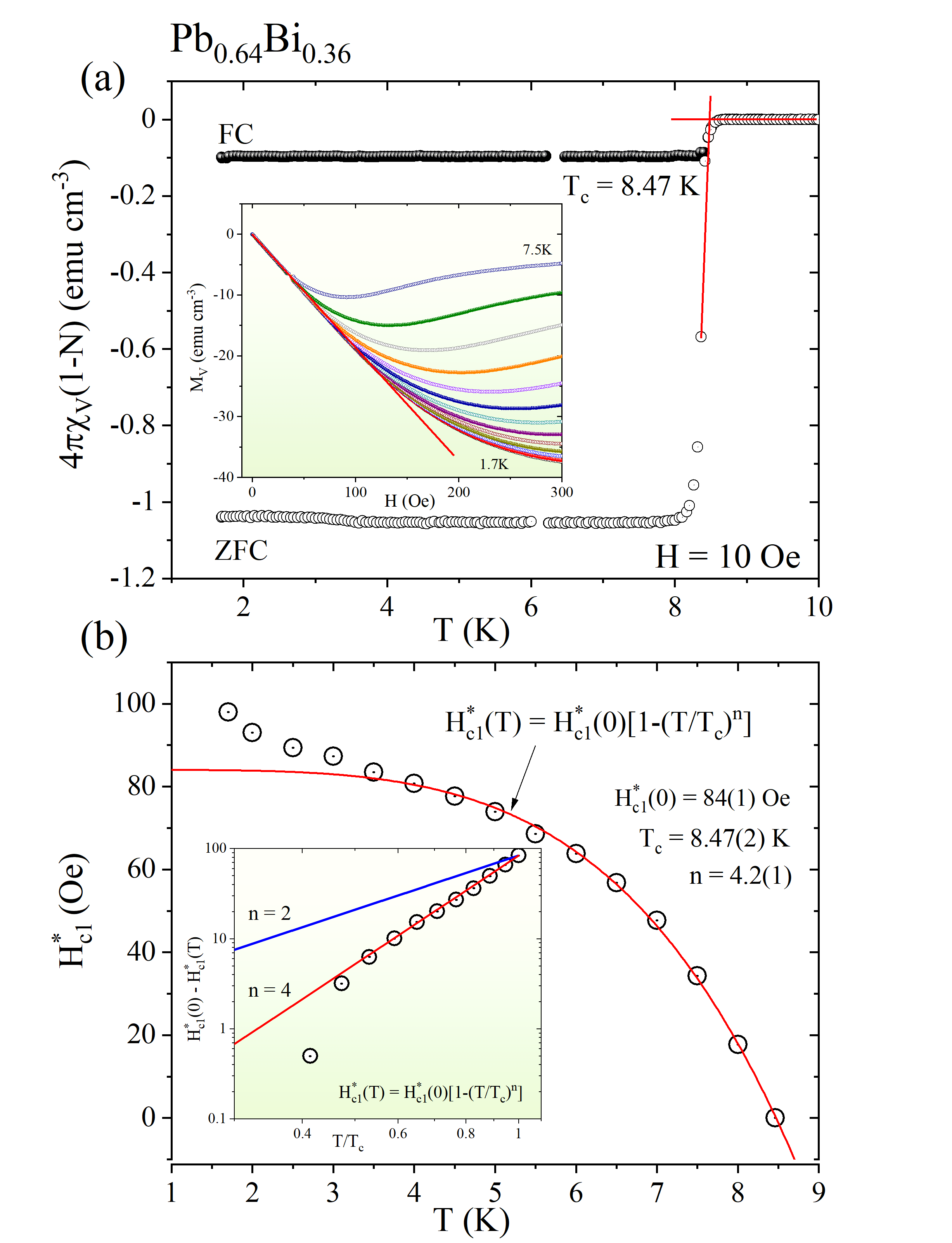}
\caption{a)  Zero-field-cooled (ZFC) and field-cooled (FC) temperature-dependent volume magnetic susceptibility measurements performed under a small applied magnetic field of 10 Oe. The inset shows the field-dependent magnetization curves taken at different temperatures; b) Temperature variation of the lower critical field derived from the inset of panel (a). The red line represents the fit of Eq. (1) to the experimental data, where $n$ was considered as a fitting parameter; The inset presents the fit for $n=2$, $n=4$. }\label{fig:exp-magnetic}
\end{figure}

To characterize the superconducting transition of Pb$_{0.64}$Bi$_{0.36}$, zero field-cooled (ZFC) and field-cooled (FC) dc magnetic susceptibilities (defined as $\chi =M/H$ where $M$ is the magnetization and $H$ is the applied magnetic field)  measured during heating under an applied field of 10 Oe are shown in Fig. \ref{fig:exp-magnetic}(a).
The well-defined and sharp superconducting transition appears at $T_c = 8.47$ K, where the critical temperature ($T_c$) was determined as an intersection between the line obtained by extrapolation of the normal-state magnetic susceptibility to lower temperatures and the line drawn at the steepest slope of the superconducting signal in the ZFC data \cite{klimczuk2004carbon}. 
The diamagnetic susceptibility corrected for the demagnetization factor $N = 0.58$ (obtained from the volume magnetization $M_V(H)$ fit discussed next) approaches a value of -1 for $T \le 8$ K, corresponding to a $100\%$ shielding fraction. The much weaker field-cooled signal compared to the ZFC data is usually observed in polycrystalline samples. The inset of Fig. \ref{fig:exp-magnetic}(a) depicts magnetization curves, $M_V(H)$, in low applied magnetic fields measured for a range of temperatures ($1.7$ K $\leq T \leq$ $7.5$ K). Assuming that the initial linear response to the field for an isotherm taken at $T = 1.7$ K is ideally diamagnetic, the demagnetization factor $N = 0.58$ was found. The values of the lower critical field $H_{c1}^{*}(T)$ were extracted for each temperature as the point of deviation from the full Meissner effect \cite{umezawa1988anisotropy}. In the main panel of Fig.~\ref{fig:exp-magnetic}(b) the estimated values of $H_{c1}^{*}$ are presented. An additional point for $H = 0$ is the zero field transition temperature taken from the heat capacity measurement. The experimental data points were analyzed with the formula: 
\begin{equation}
    H^*_{c1}(T)=H^*_{c1}(0)\left(1- \left(\frac{T}{T_c}\right)^n\right)
\end{equation}
where $H^*_{c1}(0)$ is the lower critical field at $0$ K and $T_c$ is the superconducting critical temperature. A typical temperature dependence of the lower critical field values is quadratic ($n = 2$) although there is no fundamental significance of the parabolic character \cite{RoseInnes2019IntroductionTS}. The fit yielded the parameters $T_c = 8.47(2)$ K, $H^*_{c1}(0) = 84(1)$ Oe and $n = 4.2(1)$. When correcting for the demagnetization factor ($N = 0.58$), the lower critical field at $0$ K is calculated to be $H_{c1}(0) = 200$ Oe. There are two striking features observed in $H_{c1}^{*}(T)$ behavior. 
The estimated exponent value $n$ is twice the typical value $n = 2$. The inset of Fig. \ref{fig:exp-magnetic}(b) clearly shows that the experimental data points for $0.5 < \frac{T}{T_c} < 1$ follow a red line that represents the $n = 4$ model. A deviation from the parabolic shape was also observed for strongly-coupled superconductors \{Nb,Ta\}Ir$_2$B$_2$ \cite{gornicka2021nbir2b2}. 
Another interesting feature is seen below $3.5$ K. The data points for $T < 3.5$ K diverge from the model. 
These features indicate a possibility for an anisotropic and multigap nature of the superconducting state.
We would like to point out that this effect was observed for other tested samples with Pb:Bi ratio 0.62:0.38 and 0.66:0.34.

 \begin{figure}[b]
\includegraphics[width=.47\textwidth]{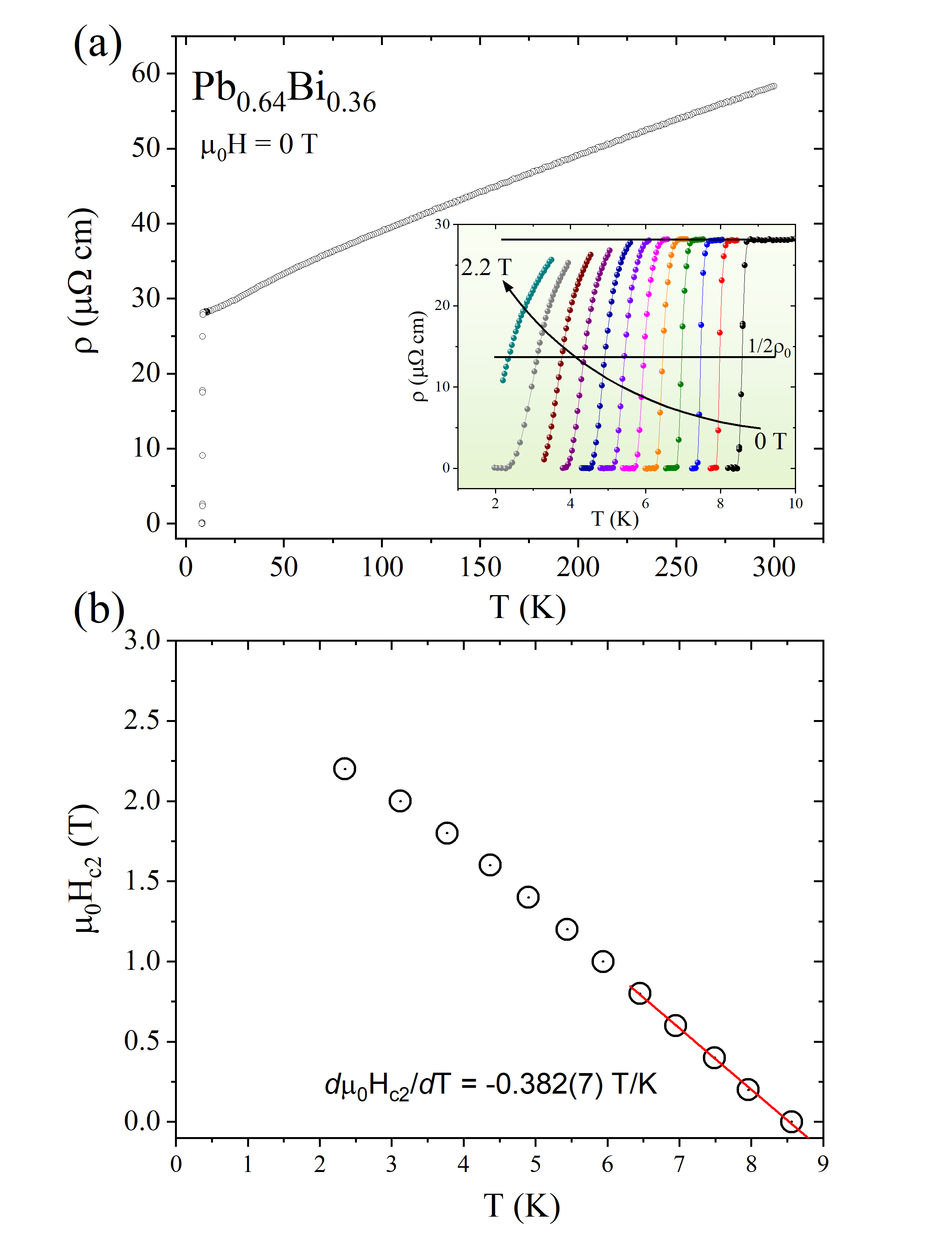}
\caption{a) The temperature dependence of the electrical resistivity in zero magnetic field. Inset: the magnetic field-dependent resistivity in the vicinity of the superconducting transition.  b) The temperature dependence of the upper critical field  determined from electrical resistivity measurements. The solid straight line represents the initial slope of the upper critical field curve }\label{fig:exp-resistivity}
\end{figure}

The main panel of Fig. \ref{fig:exp-resistivity}(a) presents the results of electrical resistivity measurements $\rho(T)$ for temperatures up to 300 K.
In the normal state, resistivity decreases with temperature, indicating metallic behavior ($\frac{d\rho}{dT}> 0$). The residual resistivity ratio $RRR = \frac{\rho(300~{\rm K})}{\rho(9~{\rm K})} = 2.1$ is rather low. 
At lower temperatures, the curve $\rho(T)$ undergoes a sudden drop at $T_c = 8.60$ K, indicating the transition to the superconducting state. For further analysis, the critical temperatures are defined as $50\%$ decrease in resistivity with respect to its normal state value ($\rho_0$). 
The inset of Fig. \ref{fig:exp-resistivity}(a) emphasizes the low-temperature resistivity under various magnetic fields from $0$ to $2.2$ T. As the magnetic field increases, the transition becomes slightly broader and $T_c$ shifts to a lower temperature. Using the same criterion to obtain $T_c$ for different applied magnetic fields, the upper critical fields are plotted as a function of temperature in Fig. \ref{fig:exp-resistivity}(b). Using the Werthamer-Helfand-Hohenberg relation \cite{werthamer1966temperature,helfand1966temperature} the upper critical field $\mu_0H_{c2}(0)$ can be estimated from
\begin{equation}
\mu_0 H_{c2}(0)=-AT_c \frac{d\mu_0 H_{c2}}{dT}|_{T=T_c}    
\end{equation}
where A = 0.69 is the purity factor for the dirty limit, expected to hold for this alloy.
We used the WHH formalism by taking the slope in the low-field region, where the data follow a significant linear dependence. A linear fit to the data gave the coefficient $\frac{d\mu_0H_{c2}}{dT} = – 0.382(7)$~T/K and taking $T_c$ = 8.47 K, the upper critical field is $\mu_0H_{c2} = 2.23$ T. Consequently, the coherence length, $\xi_{GL}$, can be estimated using the Ginzburg-Landau formula $H_{c2}=\frac{\Phi_0}{2\pi\xi^2_{GL}}$, where $\phi_0 = \frac{hc}{2e}$ is the quantum flux. For $\mu_0H_{c2}(0) = 2.23$ T, we obtain $\xi_{GL}(0) = 118$ \AA. 
The Ginzburg-Landau superconducting penetration depth $\lambda_{GL}$, calculated using the relation $H_{c1}=\frac{\Phi_0}{4\pi\lambda^2_{GL}\ln\frac{\lambda_{GL}}{\xi_{GL}}}$ is $1460$ \AA. Furthermore, the Ginzburg-Landau parameter $\kappa_{GL}= \frac{\lambda_{GL}}{\xi_{GL}}$ is estimated to be $\kappa_{GL}= 13$, which is larger than the limiting value of $\frac{1}{\sqrt{2}}$ for type-I superconductors, confirming that Pb$_{0.64}$Bi$_{0.36}$ is a type-II superconductor. The value of the thermodynamic critical field, which provides a measure of the superconducting condensation energy, can be obtained from $\kappa_{GL}$, H$_{c1}$ and H$_{c2}$ using the formula $H_{c1}H_{c2}=H_c^2\ln \kappa_{GL}$. The resulting value of $H_c(0)$ (from the $H_{c1,2}$ extrapolated to 0~K) is $1350$ Oe. $H_c$ was also calculated as a function of temperature 
and will be shown together with theoretical calculations in the next Section.

\begin{figure}[b]
\includegraphics[width=.47\textwidth]{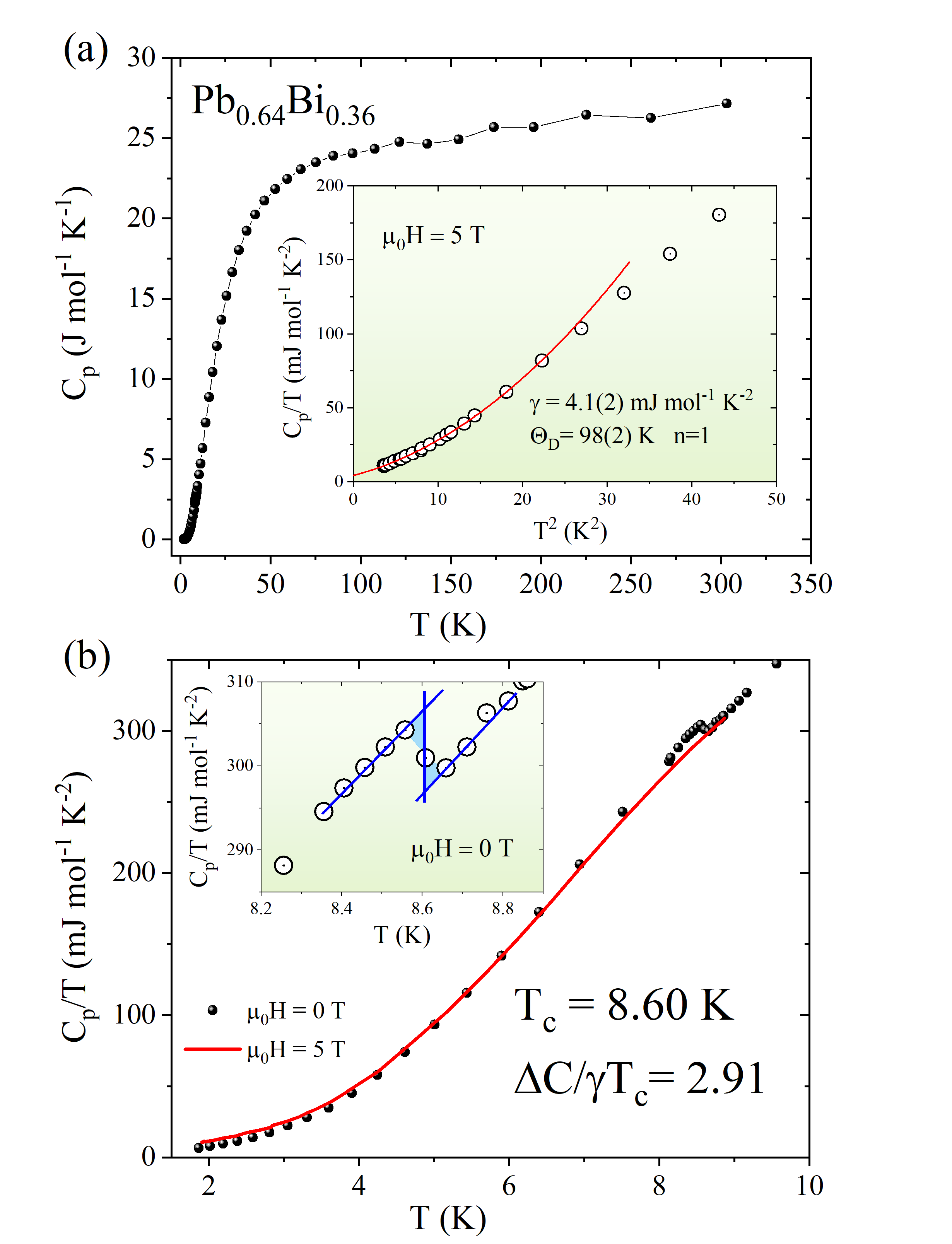}
\caption{a) The specific heat measured from 1.95 to 300 K under zero magnetic field. The inset shows $C_p/T$ data versus $T^2$. b) Specific heat data in the vicinity of the superconducting transition measured without (black dots) and with an applied magnetic field of 5 T (red solid line). The inset presents the specific heat anomaly in a zero magnetic field. The thin solid lines illustrate the equal-entropy construction used to derive the critical temperature.  }\label{fig:exp-heat}
\end{figure}

To further characterize the superconductivity of Pb$_{0.64}$Bi$_{0.36}$, temperature-dependent specific heat measurements were performed at $0$ T and a magnetic field of $5$ T, and the resulting data are presented in Fig.~\ref{fig:exp-heat}. The main panel of Fig.~\ref{fig:exp-heat}(a) shows the temperature dependence of the zero-field specific heat $C_p$ from $1.95$ to $300$ K. 
At high temperature, $C_p$ reaches the expected value calculated from the Dulong-Petit law $3nR \simeq 25~\frac{\rm J}{{\rm mol}\cdot {\rm K}}$, where $n = 1$ is the number of atoms per formula unit and $R = 8.314~\frac{\rm J}{{\rm mol}\cdot {\rm K}}$ is the gas constant.  The inset of Fig. \ref{fig:exp-heat}(a) shows the heat capacity data plotted as $C_p/T$ versus $T^2$, under the magnetic field of $5$ T. Our data were fitted using equation $C_p/T = \gamma + \beta T^2 + \delta T^4$, where the first term is the electronic specific heat coefficient and the second and third terms are attributed to the contribution of the lattice to the heat capacity. 
From the fit shown by the solid red line we obtained $\gamma = 4.1(2)~\frac{\rm mJ}{{\rm mol}\cdot {\rm K}^2}$, $\beta = 1.47(1) ~\frac{\rm mJ}{{\rm mol}\cdot {\rm K}^4}$ and $\delta = 0.101(8)~\frac{\rm mJ}{{\rm mol}\cdot {\rm K}^6}$. The Debye temperature $\Theta_D$ is then calculated from the $\beta$ coefficient, $\Theta_D=\Big(\frac{12\pi^4}{5\beta}nR\Big)^{\frac{1}{3}}$, where R is a gas constant and 
$n = 1$. Using the value of $\beta$, we obtained the value of $\Theta_D = 98(2)$ K, which is between the Debye temperatures for Pb ($\Theta_D= 95$ K) and Bi ($\Theta_D=115$ K). 
Having the estimated Debye temperature $\Theta_D$, the electron-phonon constant $\lambda$ is usually calculated from the inverted McMillan formula~\cite{mcmillan}:
\begin{equation}\label{eq:mcmillan}
\lambda = \frac{1.04+\mu^*\ln\Big(\frac{\Theta_D}{1.45T_c}\Big)}
{(1-0.62\mu^*)\ln\Big(\frac{\Theta_D}{1.45T_c}\Big)-1.04}
\end{equation}
where $\mu^*$ is the retarded Coulomb pseudopotential parameter, typically taken as $\mu^* = 0.10 - 0.13$ \cite{mcmillan}. Taking $T_c = 8.60$ K and $\Theta_D = 98$ K, we obtained $\lambda = 1.40 - 1.53$, indicating that Pb$_{0.64}$Bi$_{0.36}$ may be a strong-coupling superconductor.

\begin{table}[b]
	\caption{Summary of superconducting properties of  Pb$_{0.64}$Bi$_{0.36}$ determined in this work, data for Pb are taken from Refs.~\cite{pb-hc,pb-heat1,pb-heat2,pb-heat3}.	}					
	\label{tab:expt}	    
    \centering
\begin{ruledtabular}
    \begin{tabular}{lc c c r}
Parameter	&	Unit	& Pb$_{0.64}$Bi$_{0.36}$  &Pb \\						
	\hline									
$T_c$	&	K	&	8.6	&	7.2	\\			
$\mu_0H_{c1}(0)$	&	Oe	&	200	&	802	\\			
$\mu_0H_{c2}(0)$	&	T	&	2.23	&	\\				
$\mu_0H_{c}(0)$	&	Oe	&	1350		\\				
$\lambda_{e-p}$	(Eq. \ref{eq:tc-modified}) &	-	&	2.09		& 1.6	\\			
$\xi_{GL}(0)$	&	\AA	&	118		\\				
$\lambda_{GL}(0)$	&	\AA	&	1460		\\				
$\kappa_{GL}$	&	-	&	13		\\				
$\gamma$	&	$\frac{\rm	mJ}{{\rm	mol}	\cdot	{\rm	K}^2}$	&	4.1(2)	& 3.0--3.1\\
$\beta$	&	$\frac{\rm	mJ}{{\rm	mol}	\cdot	{\rm	K}^4}$	&	1.47(1)	&\\
$\theta_D$	&	K	&						98(2)	&96--105\\
$\frac{\Delta	C_p}{\gamma	T_c}$	&	-	&	2.9	&2.68			\\
$RRR$	&	-	&	2.1						\\
	\end{tabular}																
\end{ruledtabular}
\end{table}										

The main panel of Fig.~\ref{fig:exp-heat}(b) shows the heat capacity data at low temperatures measured in the vicinity of the transition under the zero and $5$~T magnetic field. The inset shows the zero-field data plotted as $C_p/T$ versus $T$. The jump visible in the specific heat data at $T_c = 8.60$ K suggests a good quality of the sample and confirms the bulk superconductivity in the material. It is known that the normalized specific heat jump, $\frac{\Delta C}{\gamma T_c}$, can be used to measure the strength of the electron coupling. In the BCS weak-coupling limit, its value is $1.43$. For Pb$_{0.64}$Bi$_{0.36}$, the ratio of $\frac{\Delta C}{\gamma T_c}$ estimated from the above data is approximately $2.9$, significantly higher than the weak coupling value. This indicates that Pb$_{0.64}$Bi$_{0.36}$ is a strong-coupling superconductor. 

In such a case of strong EPC, the McMillan equation is less accurate. 
Instead, the Allen-Dynes formula for $T_c$ \cite{allen-dynes} should be used. Before, 
the logarithmic average phonon frequency $\omega_{\rm log}^{\alpha^2F}$ has to be calculated from the equation for the specific heat jump in strongly-coupled superconductors~\cite{marsiglio-carbotte}: 
\begin{equation}\label{eq:marsiglio}
\frac{\Delta C_p}{\gamma T_c}=1.43\left[1+53\left(\frac{T_c}{\omega_{\rm log}^{\alpha^2F}}\right)^2\ln\left(\frac{\omega_{\rm log}^{\alpha^2F}}{3T_c}\right)\right].
\end{equation}
With $T_c=8.60$ K and $\frac{\Delta C_p}{\gamma T_c}=2.9$ we obtain $\omega_{\rm log}^{\alpha^2F} = 51$\,K. 

The Allen-Dynes formula with corrections for the strong coupling reads \cite{allen-dynes}
\begin{equation}
T_c=f_1f_2\frac{\omega_{\rm log}^{\alpha^2F}}{1.2}
\exp\left[ 
\frac{-1.04(1+\lambda)}{\lambda-\mu^*(1+0.62\lambda)}\right],
\label{eq:tc-modified}
\end{equation}
where
\begin{equation}
f_1=\Big(1+\big(\frac{\lambda}{\Lambda_1}\big)^{3/2}\Big)^{1/3}
\label{eq:tc-f1}
\end{equation}
\begin{equation}
f_2=1+ \frac{\Big(\frac{\sqrt{\hat{\omega}^2}}{\omega_{\rm log}^{\alpha^2F}}-1\Big) \lambda^2}{\lambda^2+\Lambda^2}
\label{eq:tc-f2}
\end{equation}
with $\Lambda_1=2.46(1+3.8\mu^*)$ and  $\Lambda_2=1.82(1+6.3\mu^*)\sqrt{\hat{\omega}^2}/\omega_{\rm log}^{\alpha^2F}$. The frequency moment appeared in this equation, 
\begin{equation}
\hat{\omega}^2=\frac{2}{\lambda}\int_0^{\omega_{max}} \omega \cdot  \alpha^2F(\omega) d\omega
\label{eq:omega2}
\end{equation}
 requires the Eliashberg function $\alpha^2F(\omega)$ to be calculated. It is done later in the text, while now we will approximately use: $\sqrt{\hat{\omega}^2}\simeq \omega_{\rm log}^{\alpha^2F}=51$ K.
In this way, by solving Eqs. (\ref{eq:tc-modified}-\ref{eq:tc-f2}) we get $\lambda=2.09$ for $\mu^* = 0.10$. 
This result shows the strong coupling character of superconductivity in the Pb-Bi alloy and is in good agreement with the value $\lambda=2.13$ obtained for Pb$_{0.65}$Bi$_{0.35}$ from the tunneling data \cite{allen-dynes,pbbi-history5}.  

Our experimental studies are summarized in Table~\ref{tab:expt}. The superconducting state parameters show a significant enhancement compared to elemental Pb, and the evolution of the fundamental properties of the system upon alloying with Bi is explained using the theoretical calculations in the next part of this work.

\section{Theoretical studies}

\subsection{Crystal structure}

\begin{figure}[b]
\includegraphics[width=0.46\textwidth]{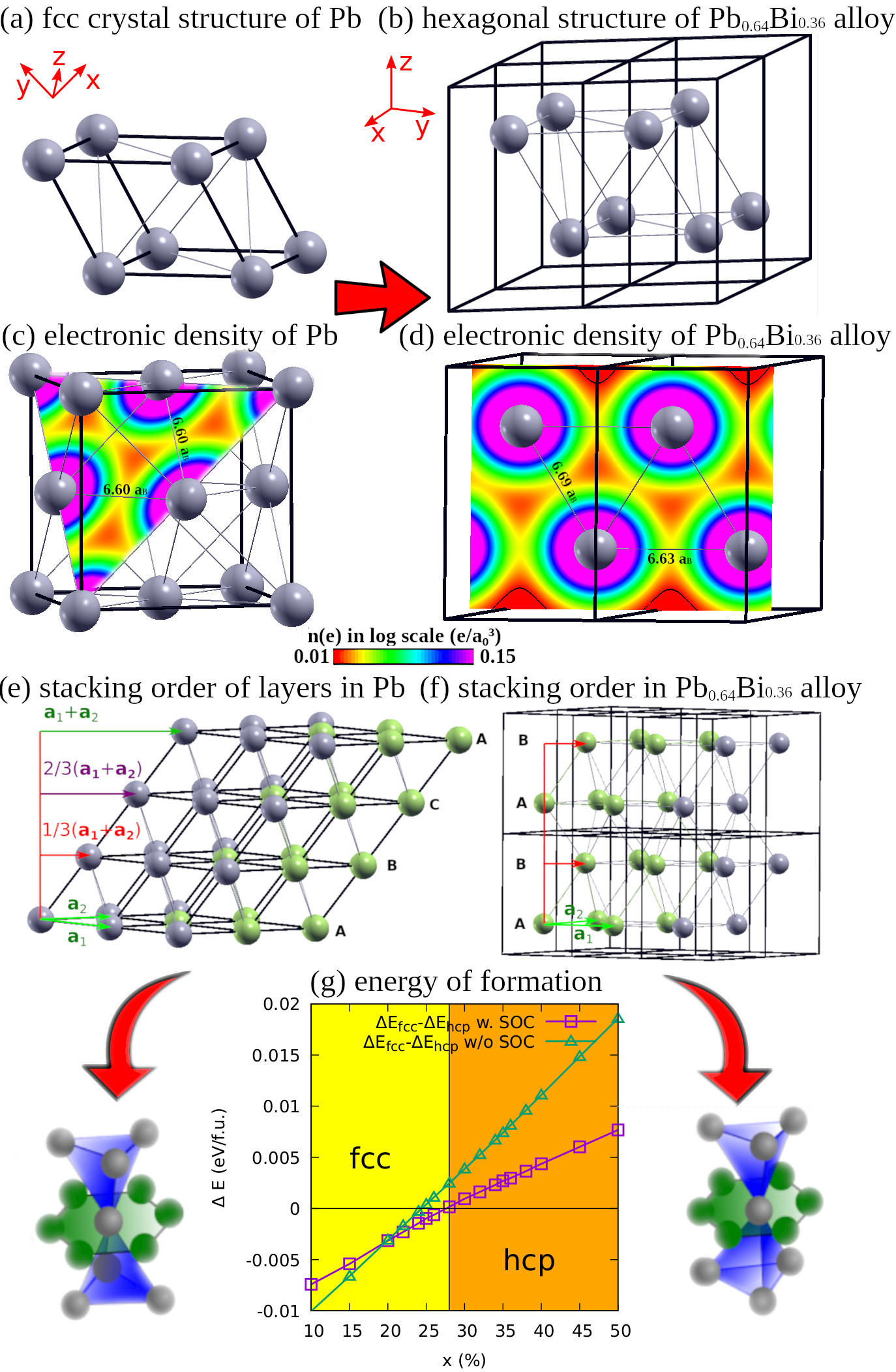}
\caption{The crystal structures of (a) $fcc$ Pb and (b) hexagonal (nearly $hcp$) Pb$_{0.64}$Bi$_{0.36}$ together with their charge densities shown in  (c) (111) and  (d) (100) planes, respectively. The hexagonal structure is presented in 2x2x1 supercell to underline its similarity to $fcc$ structure. The two structures are made of the same hexagonal atomic layers stacked in [111] and [001] directions respectively, shifted with a vector $(\frac{1}{3},\frac{1}{3},0)$ (in crystal coordinates) with respect to each other. The different stacking of these layers (A-B-C-A-... in $fcc$ and A-B-A-... in $hcp$), shown in panels (e) and (f) leads to a different (cubic and hexagonal) crystal symmetry. On the sides of panel (g) the nearest neighbors of atom in $fcc$ and $hcp$ structures are presented. The  difference of formation energy of $fcc$ and hexagonal phases of Pb$_{1-x}$Bi$_x$ is shown in panel (g). The formation energy is calculated as $\Delta E_{\rm fcc/hcp}=E_{\rm fcc/hcp}-E_{\rm Bi}-E_{\rm Pb}$, where $E_{\rm fcc/hcp}$, $E_{\rm Bi}$ and $E_{\rm Pb}$ are the total energies of Pb-Bi alloy, Bi in trigonal structure and Pb in $fcc$ structure respectively. All of considered structures where relaxed and calculated in two versions - without and with SOC.}\label{fig:charge}
\end{figure}

Both the $fcc$ structure of Pb and the hexagonal structure of Pb$_{0.64}$Bi$_{0.36}$ are shown in Fig. \ref{fig:charge}(a-b), where the latter is shown in a 2x2x1 supercell to highlight its relationship to the $fcc$ structure. 
As is commonly known \cite{kittel1955solid}, in the $hcp$ structure, the hexagonal $xy$ layers of the atoms are equivalent to the (111) layers of the $fcc$ cell; however, their stacking is different, leading to a different symmetry (see Fig. \ref{fig:charge}(e-f)). 
In the elementary $fcc$ cell, every atom has 12 nearest neighbors (NN), as shown on the left side of Fig. \ref{fig:charge}(g). The same holds for the ideal $hcp$ structure with $c/a = \sqrt{8/3} \simeq 1.63$ in spite of the different geometry, see the right side of Fig. \ref{fig:charge}(g).
However, the structure of Pb$_{1-x}$Bi$_x$ alloys is distorted along the $z$ direction and $c/a>1.63$. 
In such a structure, the $xy$ hexagonal closed-packed layers of atoms (each having six NN) are slightly separated from each other. As a result, the metallic atomic bonds, visible in the charge density shown in Fig. \ref{fig:charge}(c-d) are slightly stronger in the $xy$ plane than between planes. This will be important for the electron-phonon interaction in the system.

According to the experiment \cite{pbbi-history7} the hexagonal phase of Pb$_{1-x}$Bi$_x$  generally forms above $x=0.20$, but it co-exists with the $fcc$ phase up to $x \simeq 0.30$.
The lattice parameters $a$ and $c/a$ are about $3.50$~\AA ~and $1.65$, respectively, almost constant across the entire composition range (from 25\% to 50\% the lattice parameter $x$ changes less than $0.3\%$). 
Our calculations of the formation energy confirm the tendency to the $fcc$-hexagonal transition. The formation energy, calculated as the difference in the energy of the $fcc$/hexagonal alloy and the energy of $fcc$ Pb and trigonal Bi, is shown in Fig. \ref{fig:charge}(g).
It shows the preference for the hexagonal structure above $x=0.28$ and only small variations of the lattice parameters with $x$. Additionally, the effect of SOC slightly shifts the phase transition point, as without SOC it would happen at $x=0.24$.

In the following, we analyze how the electronic structure changes from Pb to Pb-Bi alloy, and how it is influenced by the additional electron from Bi and by the spin-orbit coupling. 

\begin{figure}[t]  
\includegraphics[width=.47\textwidth]{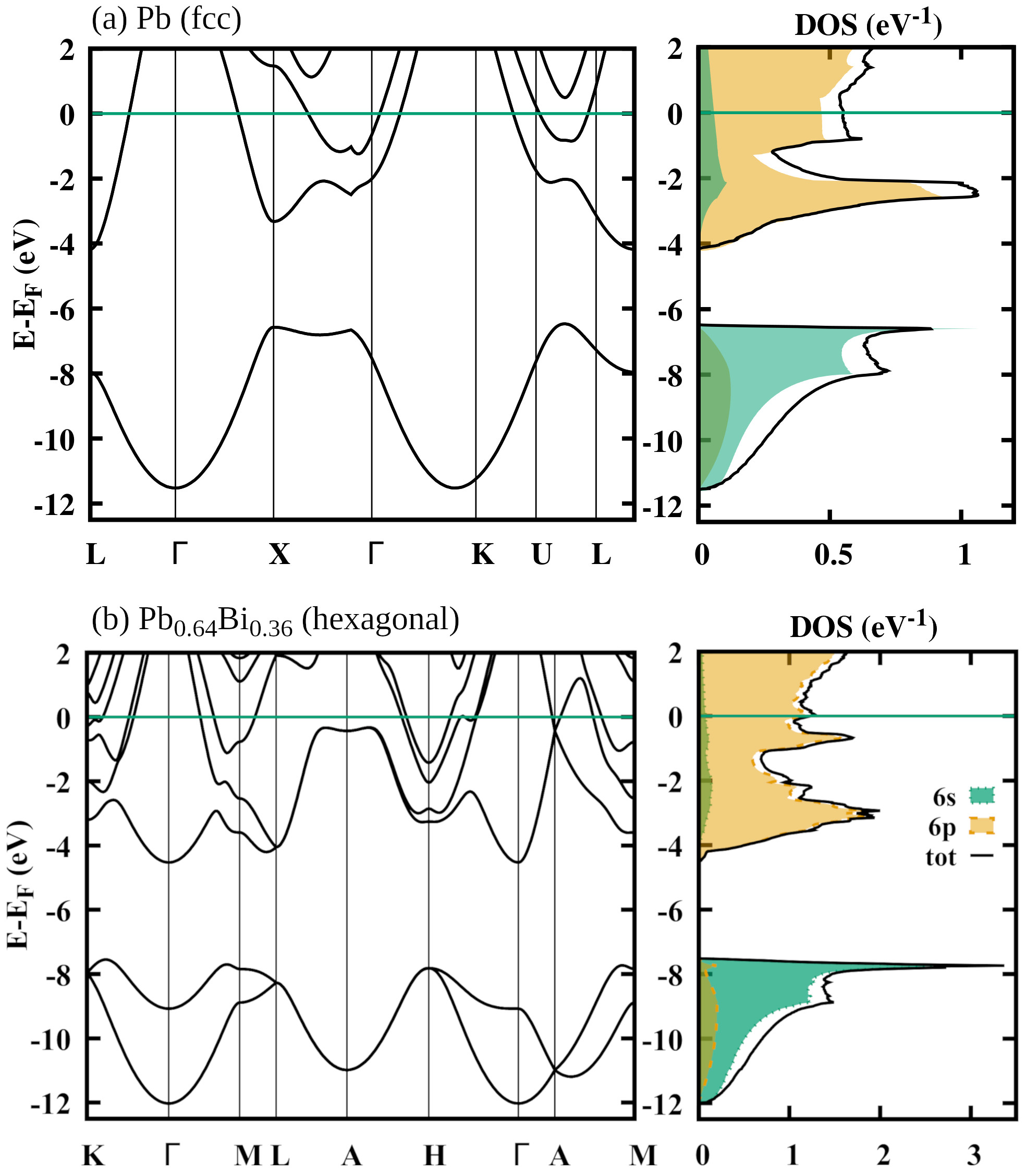}
\caption{The electronic structure of (a) $fcc$ Pb; and (b) hexagonal Pb$_{0.64}$Bi$_{0.36}$ alloy; in terms of bands and DOS, calculated with SOC. The DOS projected onto atomic orbitals is marked with colors.}
\label{fig:el}
\end{figure}

\subsection{Electronic structure}

A general view of the electronic structures of Pb and Pb$_{0.64}$Bi$_{0.36}$ is shown in Fig. \ref{fig:el}.
In both cases, the $6s$ states are located between -12 eV and -7 eV below the Fermi level.
The main valence band is formed by the $6p$ orbitals and starts 4 eV below $E_F$. 
In the case of Pb, it is occupied by two $6p$ electrons per atom, while in Pb$_{0.64}$Bi$_{0.36}$ alloy the number of occupied states increases to 2.36 per f.u. (4.72 per unit cell).

\begin{figure}[t]
\includegraphics[width=.47\textwidth]{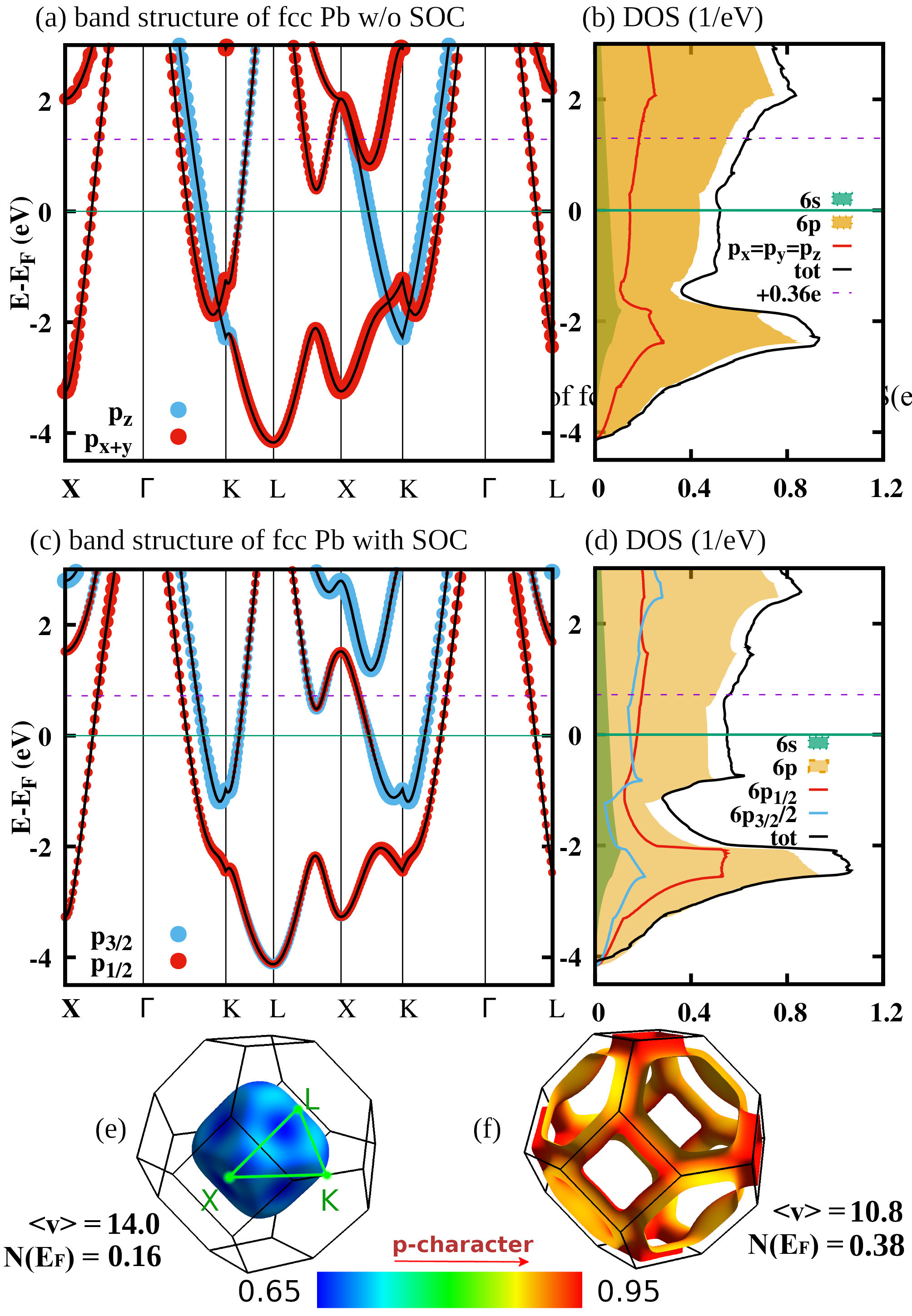}
\caption{The band structure of Pb calculated without (a-b) and with (c-d) SOC in terms of bands (a, c), DOS (b, d) and Fermi surface (e-f). The bands are colored according to their orbital character. There are two positions of the Fermi level marked on panels (a-d): one for pure Pb (solid line) and the second shifted to accommodate additional 0.36 electrons (dashed line). 
The Fermi surface is colored according to $p$-orbital contribution, between 65\% and 95\%. In panel (e) the high symmetry points of Brillouin zone are shown. Additionally, the mean value of Fermi velocity $\langle v \rangle$ and the contribution to the $N(E_F)$ of each FS sheet are given in the units of $10^5$\,$\frac{\rm m}{\rm s}$ and eV$^{-1}$, respectively.}\label{fig:elpb}
\end{figure}

 \begin{table}[b]
\caption{Experimental and calculated value of Sommerfeld coefficient  $\gamma$ ($\frac{{\rm mJ}\cdot {\rm mol}}{K^2}$), electronic density of states at $E_F$, $N(E_F)$ (eV$^{-1}$/f.u.) and EPC constant calculated with Eq. \ref{eq:gamma-lambda}. Scalar-relativistic results are labeled as "w/o SOC", relativistic have no additional label here. Experimental data for Pb comes from \cite{pb-heat1}.}
\label{tab:gammy}
\begin{center}
\begin{ruledtabular}
\begin{tabular}{ l c c c c  }
	&	$\gamma_{\rm expt}$	&	$N(E_F)$	&	$\gamma_{\rm band}$	&	$\lambda_{\gamma}$	\\
	\hline 
Pb	w/o SOC &	     	&	0.522	&	1.23	&	1.54	\\
Pb	 &	3.13    &	0.541	&	1.28	&	1.45 \\
Pb$_{0.64}$Bi$_{0.36}$ w/o SOC	&   	&	0.605	&	1.41	&	1.90	\\
Pb$_{0.64}$Bi$_{0.36}$ 	&	4.10	&	0.614 &	1.45	&	1.83	\\
Pb$_{0.64}$Bi$_{0.36}$ CPA 	&  4.10 	&	0.593	&	1.40	&	1.96	
\end{tabular}
\end{ruledtabular}
\end{center}
\end{table}
 
Figure \ref{fig:elpb} shows the main valence band of Pb in detail, with an orbital character marked by color.  
Starting from the scalar-relativistic case, there are two bands that cross the Fermi level, one of the $p_{x,y}$ and the second of the $p_z$ character. Fermi level is placed on the flat region of DOS, with $N(E_F) = 0.54$ eV$^{-1}$ per atom.
When SOC is included, many anticrossings of bands appear and the relativistic $p$ orbital is composed of one spherical $p_{1/2}$ and two non-spherical $p_{3/2}$. However, in the vicinity of the Fermi level, the scalar-relativistic and relativistic bands are very similar and the contribution of $p_{1/2}$ and $p_{3/2}$ is the same as that of $p_{x,y}$ and $p_z$, respectively.
Using the calculated $N(E_F)$, the bandstructure values of the Sommerfeld coefficient are calculated as 
\begin{equation}
    \label{eq:gamma}
\gamma_{\rm band}=\frac{\pi^2}{3}{k_B^2}N(E_F)
  \end{equation}
and collected in Table~\ref{tab:gammy}.
Taking the experimental result from \cite{pb-heat1}, the electron-phonon coupling constant is determined as a renormalization factor, 
\begin{equation}\label{eq:gamma-lambda}
\gamma_{\rm expt}=(1+\lambda_{\gamma})\gamma_{\rm band}
\end{equation}
and is equal to $\lambda_{\gamma}=1.42$, confirming the strong EPC in Pb.

Two sheets of the Fermi surface of $fcc$ Pb are shown in Fig.~\ref{fig:elpb}(e-f) with the orbital $p$ contribution indicated by color (the maximum value of 1 means 100\% $p$-type character).
The first is a pocket centered on $\Gamma$, which originates from hybridization of the $s$ orbitals with $p_{x,y}$ (or $p_{1/2}$ in the relativistic case). 
The second is tube-like centered along the boundaries of the Brillouin zone and is contributed by the $p_z$ orbital (or $p_{3/2}$ in the relativistic case). Both pieces contribute nearly equally to the total density of states at the Fermi level.
 
\begin{figure}[t!] 
\includegraphics[width=.47\textwidth]{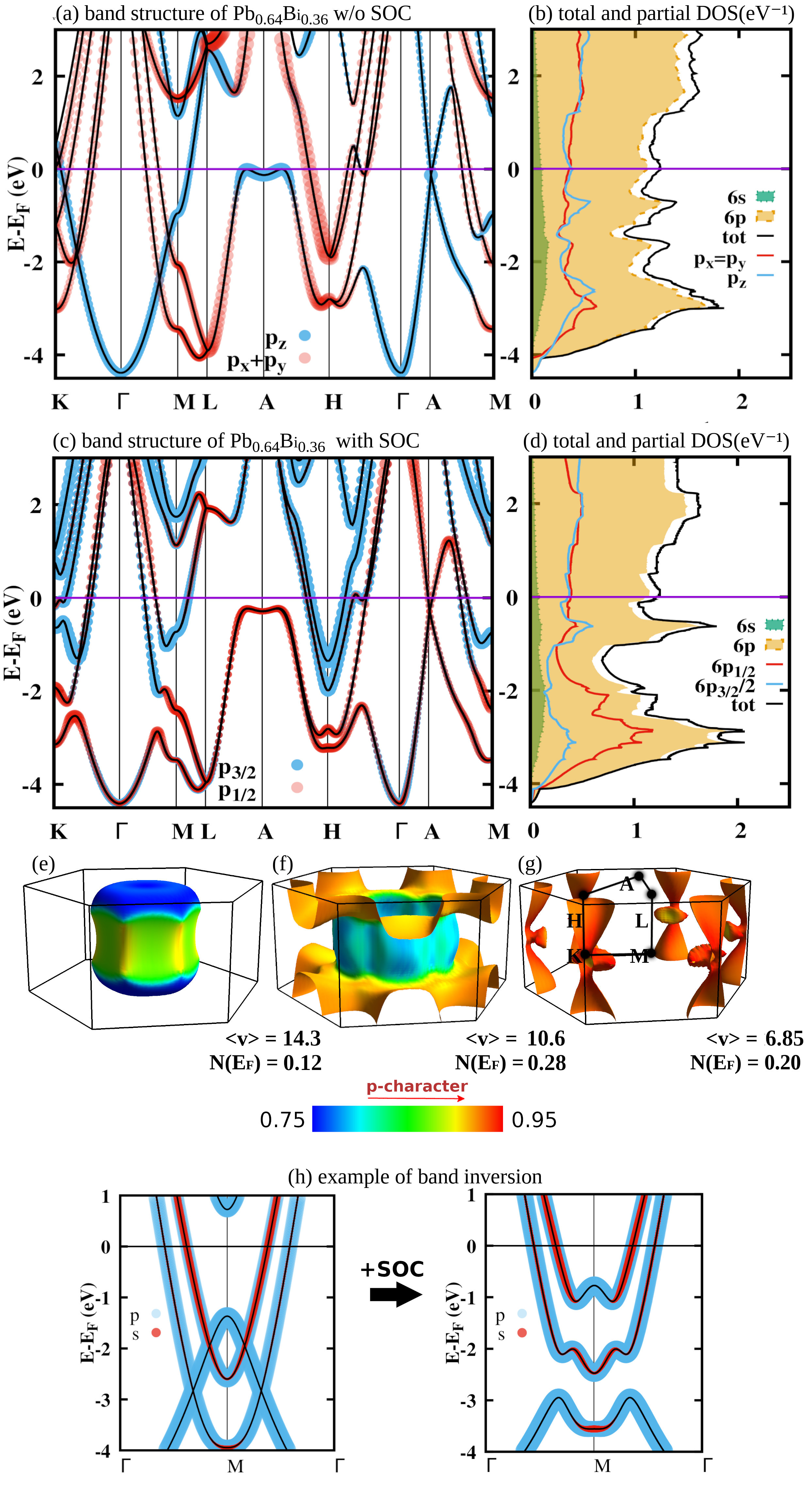}
\caption{The band structure of Pb$_{0.64}$Bi$_{0.36}$ alloy calculated without (a-b) and with (c-d) SOC in terms of bands (a, c), DOS (b, d) and Fermi surface (e-g). The bands are decorated with colored points according to their orbital character. The Fermi surface is colored with p-orbital contribution. 
At panel (g) the high symmetry points of Brillouin zone are shown. Additionally, the mean  value of Fermi velocity $\langle v \rangle$ and the contribution to the $N(E_F)$ of pieces of the Fermi surface are given in the units of $10^5$\,$\frac{\rm m}{\rm s}$ and eV$^{-1}$, respectively. In panel (h) the band structure decorated with orbital character, calculated without SOC (on the left side) and with SOC (on the right side) are compared to show the band inversion.}\label{fig:elsoc}
\end{figure}

The band structure of Pb$_{0.64}$Bi$_{0.36}$ alloy is presented in Fig. \ref{fig:elsoc} in terms of bands and DOS, calculated without SOC (panels a and b) and with SOC (panels c and d). The Fermi surface in the fully relativistic case is presented in panels (e-g).
In contrast to Pb, here the Fermi level is placed at the slope of the DOS peak. Because of that, the DOS at $E_F$ is larger in the case of alloy than in Pb, increasing to 0.614 eV$^{-1}$ per f.u. 
Comparing now the bandstructure and experimental values of the Sommerfeld parameter (see Table~\ref{tab:gammy}) the renormalization parameter increases to $\lambda_{\gamma}=1.83$, confirming a strong increase in the electron-phonon coupling strength with respect to Pb.

Similarly to the case of Pb, in Pb$_{0.64}$Bi$_{0.36}$ alloy effect of SOC leads to band shifts and anti-crossings. The effect on $N(E_F)$ is minor, it changes from $0.605$ eV$^{-1}$ per f.u. without SOC to $0.614$ eV$^{-1}$ per f.u. with SOC, see Table \ref{tab:gammy}. 
However, at some points of the Brillouin zone, band inversion is noticed. 
For example, the bands along the $\Gamma-M$ direction, presented in Fig. \ref{fig:elsoc}(h) are of $p_{3/2}$ and mixed $p_{1/2}$ and $s$ character. With SOC, the band's character, due to anticrossing of the magnitude of nearly 1 eV, becomes inverted.

In the relativistic case, three bands cross the Fermi energy, resulting in three pieces of the Fermi surface, presented in Fig. \ref{fig:elsoc}(e-g) and colored with a $p$ orbital character. 
These are: $\Gamma$-centered pocket (panel e), large sheet with a central, nearly cylindrical part (panel f), and the piece made of tubes along $z$ at the boundaries of the Brillouin zone.  
The first and third pieces are similar to those observed in Pb in Fig.~\ref{fig:elpb}. However, the orbital character of the first one is not as isotropic as that of Pb. It is caused by the above-mentioned band inversion between bands corresponding to the first and second pieces of the Fermi surface. The largest contribution to the total density of states at the Fermi level comes from the second FS sheet, which has a significant $s-p$ mixed orbital character.  
The shape of the Fermi surface of Pb$_{0.64}$Bi$_{0.36}$ reflects the distortion from cubic to layered hexagonal structure, with axial symmetry.

\begin{figure}[t]
    \includegraphics[width=.47\textwidth]{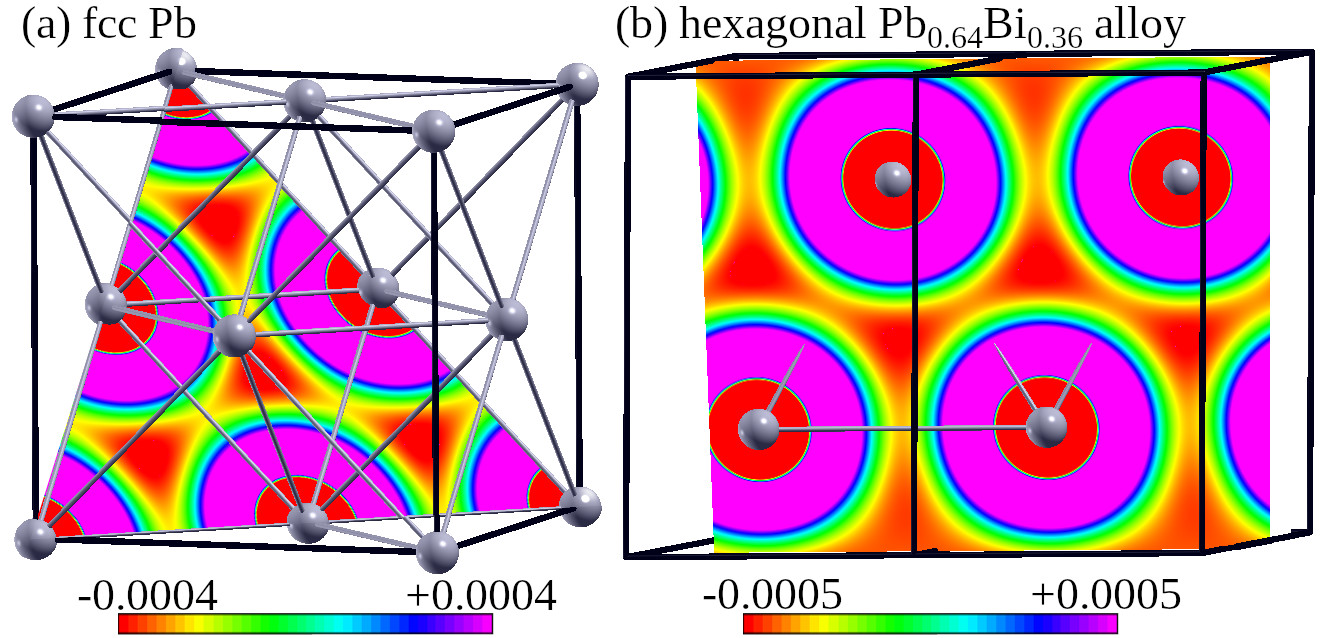}
    \caption{Difference of pseudopotential electronic charge density calculated with and without SOC, $\Delta n(\bm r)= n_{\rm SOC}(\bm r) - n_{\rm NO~SOC}(\bm r) $ ($e/a_0^3$), for a) Pb and b) Pb$_{0.64}$Bi$_{0.36}$ alloy. Due to relativistic contraction of the $p$-orbital, SOC reduces the charge density in the interstitial region.}
    \label{fig:charge-soc}
\end{figure}

The spin-orbit coupling has an important effect on the valence charge distribution. The difference in electronic pseudo-charge density calculated with and without SOC is presented in Fig. \ref{fig:charge-soc}. 
For both Pb and Pb-Bi, the charge density is reduced in the interstitial space between the atoms due to SOC.
This is caused by the relativistic $p$-orbital contraction and leads to weaker atomic bonding as a consequence of a smaller electron density in the interstitial region. This affects the phonon structure, as discussed below. 

\begin{figure}[t]
\includegraphics[width=.47\textwidth]{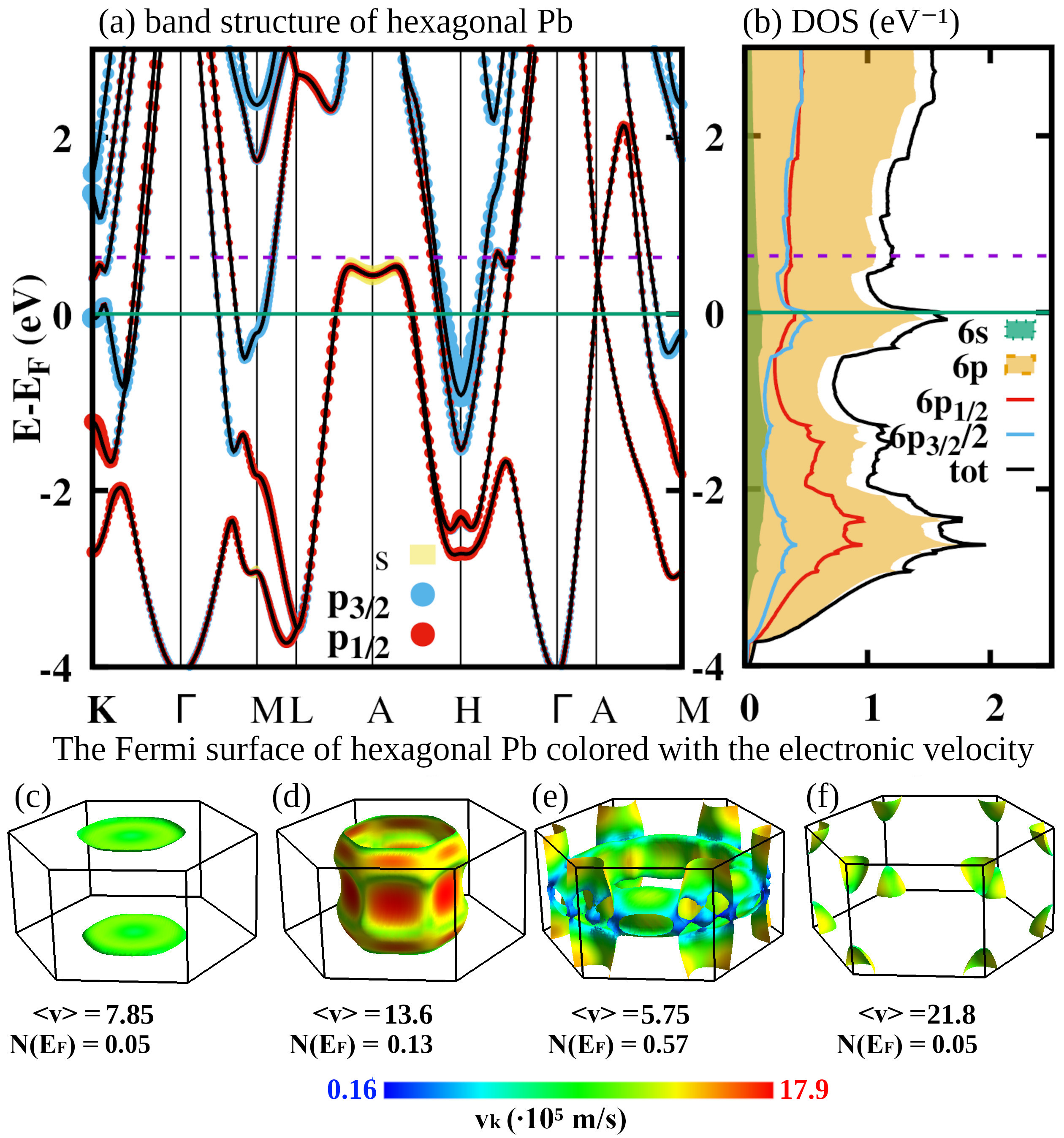}
\caption{The band structure of Pb calculated in hexagonal crystal structure of Pb$_{0.64}$Bi$_{0.36}$ alloy in terms of bands (a), DOS (b) and (c-f) Fermi surface colored with Fermi velocity ${\bf v}_{n{\bf k}}$. the Fermi level is marked with a green continuous line. The additional purple dashed line represents the Fermi level shifted so that the electron count matches that of Pb$_{0.64}$Bi$_{0.36}$ alloy.}
\label{fig:elhcppb}
\end{figure}

To understand the role of cubic-to-hexagonal transition in shaping the extraordinary properties of Pb$_{0.64}$Bi$_{0.36}$, we have calculated the electronic structure of Pb in the hexagonal structure of Pb$_{0.64}$Bi$_{0.36}$ alloy (in further text we will refer to it as {\it hexagonal Pb}).
Figure~\ref{fig:elhcppb} shows its electronic structure (computed with SOC), where the Fermi level is marked with a continuous green line.
Its position makes the hexagonal structure energetically unfavorable for Pb, since $E_F$ is located within the DOS peak, dominated by the $p_{3/2}$ states. 
However, when $E_F$ is shifted to reach the electron count of Pb$_{0.64}$Bi$_{0.36}$ alloy (the additional purple dashed line), it moves to a more stable position near a smaller peak of DOS, where the $p_{1/2}$ states dominate. 
Generally, the shape of bands, DOS and Fermi surface of hexagonal Pb is very similar to the Pb$_{0.64}$Bi$_{0.36}$, only the $E_F$ is shifted. This shows that changes in the electronic structure between Pb$_{0.64}$Bi$_{0.36}$ and hexagonal Pb generally follow the rigid-band model. 

The discussion above illustrates why the hexagonal structure is not suitable for Pb. Now, going back to the cubic structure of Pb in Fig.~\ref{fig:elpb} we can do a similar analysis to understand why the hexagonal structure is preferred when the electron concentration increases when Pb is alloyed with Bi. 
In Fig.~\ref{fig:elpb}(d) in the cubic phase, the Fermi level is in the flat region of the DOS, while the shifted $E_F$ is placed on the increasing slope of $N(E)$ formed by antibonding states (the contribution of the $p_{3/2}$ states starts to rise there). 
This induces the structural transition, as it reduces the occupation of antibonding states, forming the valley in the DOS, and placing the Fermi level near the smaller peak of the DOS, where the $p_{1/2}$ bonding states dominate.

In summary, the structural transition occurs because additional Bi electrons would occupy the antibonding states in the $fcc$ structure.
This mechanism is similar to the Peierls distortion observed in Bi \cite{jones1934applications,peierls,peierls2} (see the Supplemental Material~\cite{supplemental} for additional information on the Peierls distortion in Bi).

\subsection{Validation of VCA and effects of disorder\label{sec:disorder}}

\begin{figure}[t]
\includegraphics[width=.47\textwidth]{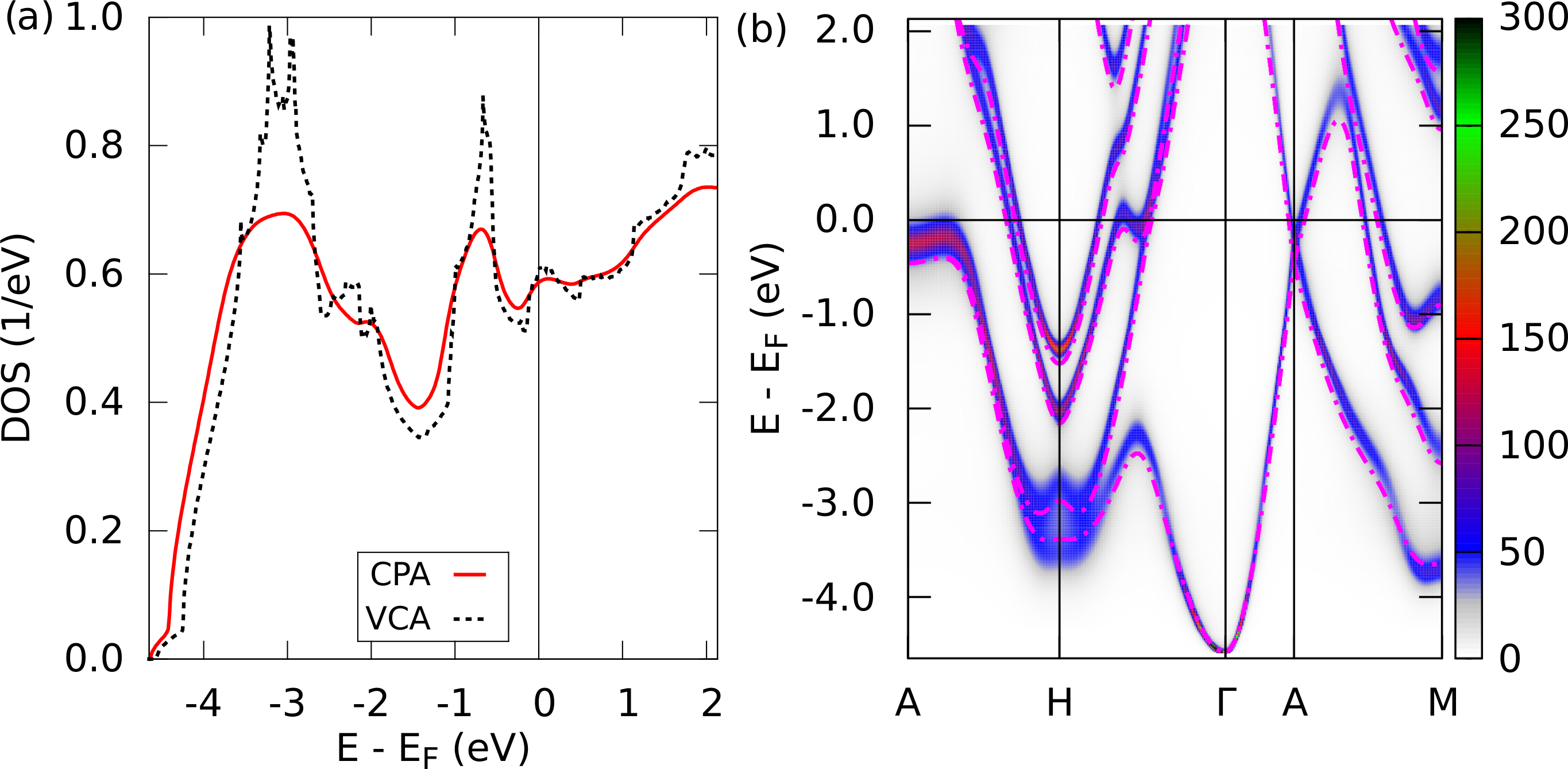}
\caption{Comparison of QE-VCA and KKR-CPA results for Pb$_{0.64}$Bi$_{0.36}$: (a) the density of states; (b) the two-dimensional projections of KKR-CPA Bloch spectral functions in high-symmetry directions (colormap in a.u.) {\it versus} VCA bands (dash-dotted magenta line). In both panels the effect of bandstructure smearing in KKR-CPA is well visible. The center of BSFs generally overlap with the VCA band positions, the peaks in DOS are smeared in KKR-CPA but the DOS near $E_F$ is similar.}
\label{fig:bsf}
\end{figure}

To validate the electronic structure of Pb$_{0.64}$Bi$_{0.36}$ calculated using the virtual crystal approximation in the mixed pseudopotential scheme, we now compare it with the results of the all-electron KKR-CPA method. 
In Fig.~\ref{fig:bsf}(a) density of states is compared. 
As characteristic for the alloy, the KKR-CPA DOS is smeared in the energies where the ordered VCA medium has peaks in DOS\footnote{The DOS was calculated using a very small imaginary part of energy of $2\times 10^{-6}$ Ry, that means the smearing of DOS peaks is not related to the integration method.}. As a consequence, the shape of DOS curves differs at lower energies, however, near $E_F$ the CPA and VCA results match. The difference in the value of $N(E_F)$ is about 4\%, KKR-CPA gives 0.593 eV$^{-1}$ per f.u., whereas the pseudopotential VCA result is 0.614 eV$^{-1}$. The lower CPA value of DOS, together with the experimental Sommerfeld parameter, result in a slightly larger electronic specific heat renormalization parameter $\lambda_{\gamma} = 1.96$,  see Table~\ref{tab:gammy}.

In Fig.~\ref{fig:bsf}(b) electronic dispersion relations are compared.
In KKR-CPA calculations for the disordered system, the electronic band structure is described using the Bloch spectral density functions (BSFs)~\cite{Ebert2011, Faulkner1980,wiendlocha2013}
$A^B(\mathbf{k},E)$, which generalize the $E({\bf k})$ dispersion relations.
BSFs are computed from the self-consistent configurationally averaged electron's Green's function.
For an ordered crystal, the BSF for each band (and spin) is a Dirac delta function, showing the position of the energy eigenvalues $E_{i,\mathbf{k}}$, thus here the BSFs define the usual dispersion relation:
\begin{equation}
A^B(\mathbf{k},E) = \sum_{i} \delta(E - E_{i,\mathbf{k}}).
\end{equation}
In the case of alloys, where a chemical disorder leads to electron scattering, the smearing of electronic bands appears and the electronic lifetime $\tau$ becomes finite. 
For typical alloys, BSF now takes the form of the Lorentz function, with the value of the full width at half-maximum (FWHM) $\Gamma_{i,\mathbf k}$ corresponding to $\tau$ \cite{gyorffy-cu_ni}:
\begin{equation}\label{eq:time1}
\tau_{i,\mathbf k} = \frac{\hbar}{\Gamma_{i,\mathbf k}}.
\end{equation}
In such a case, we can still define the energy band in alloy, with the band centers at the energy, where $A^B(\mathbf{k},E)$ has maxima, and the bandwidths corresponding to $\Gamma_{i,\mathbf k}$. This electronic lifetime $\tau_{i,\mathbf k}$ is equivalent to the Boltzmann scattering time in the Boltzmann transport equation~\cite{butler-kubo} and can be used to calculate the electrical resistivity.
The stronger the alloy scattering, the wider the BSF and the shorter the electronic scattering time. That is especially the case for the so-called resonant scattering and resonant impurities, where the BSF can lose the sharp Lorentzian shape. 
The Cu-Ni alloy is the classical example of this case~\cite{gyorffy-cu_ni,wiendlocha2018}.

\begin{figure}[t]
\includegraphics[width=.47\textwidth]{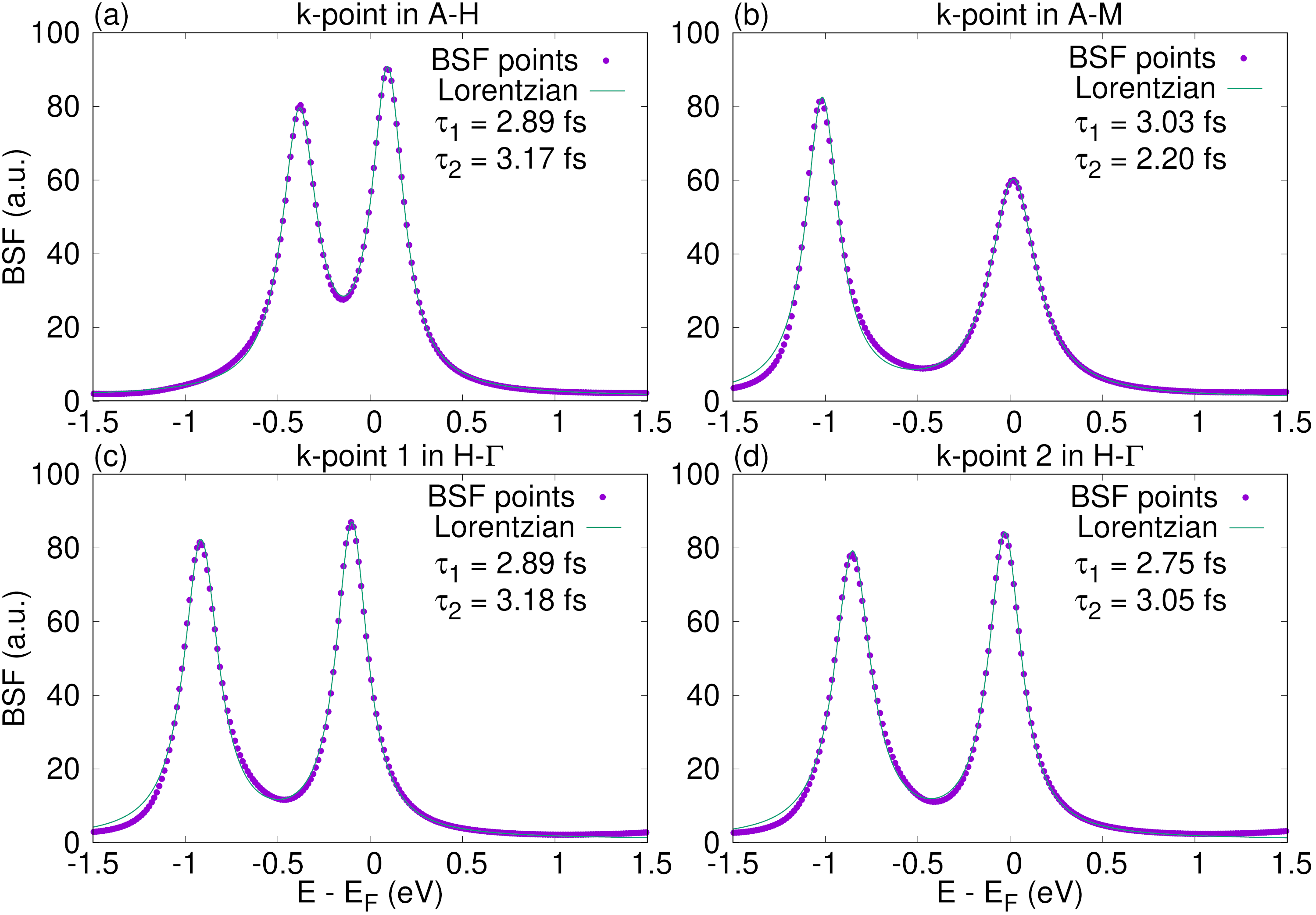}
\caption{Bloch spectral density functions at selected \textbf{k}-points where bands cross the Fermi level in the specified directions, in the energy range (-1.5, 1.5) eV. Location of \textbf{k}-points on a dispersion plot is shown in SM~\cite{supplemental}. The lifetimes $\tau$ correspond to the FWHM of the fitted Lorentz functions.}
\label{fig:bsf2}
\end{figure}

In Pb$_{0.64}$Bi$_{0.36}$ alloy, the band structure is well preserved, the spectral functions have a Lorentzian form, but with a substantial smearing.
The two-dimensional projections of the BSFs are shown in Fig.~\ref{fig:bsf}(b), where they are compared to the pseudopotential VCA bands, discussed in the previous paragraph. There are two important features in this figure. First, the position of VCA bands corresponds very well with the BSFs band centers. The largest shift between CPA and VCA is seen near the A point and is rather the effect of different computational methods (all-electron KKR-CPA vs. pseudopotentials) than related to the neglect of disorder in VCA.
Second, the spectral functions show a considerable level of smearing, showing that the alloy scattering of electrons in Pb$_{0.64}$Bi$_{0.36}$ is quite strong.
This is especially well seen in the A-H direction and around -3 eV near the H point, where we can observe a cloud of electronic states responsible for the rounded DOS at -3 eV, as seen in Fig.~\ref{fig:bsf}(a).

To analyze the scattering time, in Fig.~\ref{fig:bsf2} we have presented the spectral functions for several selected $\mathbf{k}$-points, in which bands cross the Fermi level in the A-H, A-M and H-$\Gamma$ directions. The location of these points is shown in Supplemental Material~\cite{supplemental}.
Calculated BSFs are fitted using Lorentz functions in the energy range (-1.5, 1.5) eV around $E_F$, due to the proximity of two bands a combination of two functions was used.
The scattering time $\tau$ obtained from the width of the Bloch spectral functions oscillates around $3\times 10^{-15}$ s.
This is a relatively low value; the scattering time appears here to be shorter than, e.g., in Ta-Nb-Hf-Zr-Ti high entropy alloys where it was in the range 50 - 100 fs~\cite{jasiewicz2016,jasiewicz2019,gutowska2023}. 
However, due to the relatively high Fermi velocity (average value is $v_F = 1\times 10^6$m/s), the alloy does not fall below the Mott-Ioffe-Regel limit, since the mean free path $\ell = v_F\tau \sim 30$~\AA~is much larger than the nearest-neighbor interatomic distance of 3.5~\AA, in contrast to, for example, recently investigated superconducting (ScZrNb)$_{1-x}$(RhPd)$_x$ alloys~\cite{gutowska2023}. 

A good measure of whether the alloy scattering connected to the chemical disorder is actually the main scattering mechanism in the real samples (not, e.g., the scattering on grain boundaries or dislocations) is an analysis of the residual resistivity. 
KKR-CPA calculations within the Kubo formalism gave the value of 27.2~$\mu\Omega$cm, which is in a very good agreement with the measured value of $\rho_0 = 28.2$~$\mu\Omega$cm [see Fig.~\ref{fig:exp-resistivity}]. 

The analysis of the residual resistivity and scattering time is another good test of the consistency between the VCA and CPA methods.
The absence of resonant-scattering features, which would manifest in the non-Lorentzian spectral functions, allows us to use the relaxation-time approximation and the Boltzmann transport theory in the analysis of resistivity.
Using the VCA band structure and the Boltzmann formalism in the constant relaxation time approximation, with the help of the {\sc boltztrap} code~\cite{boltztrap}
we have determined the kinetic part of the electrical conductivity tensor (that is, conductivity over the relaxation time). 
The average conductivity value obtained at $E_F$ is $\sigma/\tau = 12.0\times 10^{20}$~$\Omega^{-1}$m$^{-1}$s$^{-1}$. 
This, together with the measured residual resistivity, allows one to independently estimate the scattering time $\tau=2.75\times 10^{-15}$~s, which confirms the KKR-CPA results for $\tau$ obtained from the spectral functions.

In summary, KKR-CPA and VCA methods give consistent results, so that VCA may be used to calculate phonons and investigate the electron-phonon interaction, because electron scattering is not a dominating factor for the electronic structure. However, as the electronic relaxation time is rather short, disorder will have an impact on superconductivity, as we shall discuss later.

\subsection{Phonons \& electron-phonon coupling\label{sect:elph}}
In the spirit of the question of what makes the hexagonal Pb$_{0.64}$Bi$_{0.36}$ alloy such a strong-coupling superconductor, the phonon spectrum and electron-phonon interactions are discussed. 
To analyze the role of the three main factors: (i) the $fcc$-hexagonal structural change; (ii) Bi electron doping effect; (iii) the effect of SOC, we will analyze the phonon structure and its evolution starting from $fcc$ Pb, 
through Pb calculated in the hexagonal structure of Pb$_{0.64}$Bi$_{0.36}$ alloy
to finish at Pb$_{0.64}$Bi$_{0.36}$ system. 
Furthermore, the electron-phonon characteristics of Pb$_{0.64}$Bi$_{0.36}$ obtained without SOC are discussed to capture the influence of the spin-orbit interaction.

To characterize the phonon spectra, several phonon frequency moments are calculated
using the following formulas:
\begin{equation}\label{eq:mom}
\langle \omega^n \rangle = \int_0^{\omega_{\mathsf{max}}} \omega^{n-1} F(\omega) d\omega \left/ \int_0^{\omega_{\mathsf{max}}} F(\omega) \frac{d\omega}{{\omega}} \right.,
\end{equation}
\begin{equation}\label{eq:sred}
\langle \omega \rangle = \int_0^{\omega_{\mathsf{max}}} \omega F(\omega) d\omega \left/ \int_0^{\omega_{\mathsf{max}}} F(\omega) d\omega \right.,
\end{equation}
\begin{equation}\label{eq:omlog}
\langle\omega_{\rm log}\rangle = \exp\left(\int_0^{\omega_{\mathsf{max}}} F(\omega) \ln\omega\frac{d\omega}{{\omega}} \left/ \int_0^{\omega_{\mathsf{max}}} 
{F(\omega)}\frac{d\omega}{{\omega}} \right. \right).
\end{equation}

On the basis of phonon spectrum, the electron-phonon coupling matrix elements $g_{{\bf q}\nu}({\bf k},i,j)$ are  calculated as \cite{grimvall,wierzbowska,heid-pb,gustino-rmp}
\begin{equation}\begin{split}
g_{{\bf q}\nu}({\bf k},i,j)& =\\
=\sum_s &\sqrt{{\frac{\hbar}{ 2M_s\omega_{{\bf q}\nu}}}}
\langle\psi_{i,{\bf k+q}}| { \frac{dV_{\rm SCF}}{d {\hat u}_{\nu s}} }\cdot
                   \hat \epsilon_{\nu s}|\psi_{j,{\bf k}}\rangle,
\label{eq:el-ph-matrix}
\end{split}\end{equation}
where  $i,j$ are band indexes, $M_s$ is a mass of atom $s$, $ \frac{dV_{\rm SCF}}{d {\hat u}_{\nu s}}$ is a change of electronic potential calculated in self-consistent cycle due to the movement of an atom $s$, $\hat{\epsilon}_{\nu s}$ is a polarization vector associated with $\nu$-th phonon mode ${\hat u}_{\nu s}$ and $\psi_{i,{\bf k}}$ is the electronic wave function.
On this basis, the phonon linewidths $\gamma_{{\bf q}\nu}$ are calculated by summing $g_{{\bf q}\nu}({\bf k},i,j)$ over all the electronic states on the Fermi surface, which may interact with the given phonon $\{{\bf q}\nu\}$ \cite{grimvall,wierzbowska,heid-pb,gustino-rmp}:
\begin{equation}
\begin{split}
\gamma_{{\bf q}\nu} =& 2\pi\omega_{{\bf q}\nu} \sum_{ij}
                \int {\frac{d^3k}{ \Omega_{\rm BZ}}}  |g_{{\bf q}\nu}({\bf k},i,j)|^2 \\
                    &\times\delta(E_{i,{\bf k}} - E_F)  \delta(E_{j,{\bf k+q}} - E_F).
\end{split}
\end{equation}

\begin{figure*}[htb!]
\includegraphics[width=0.99\textwidth]{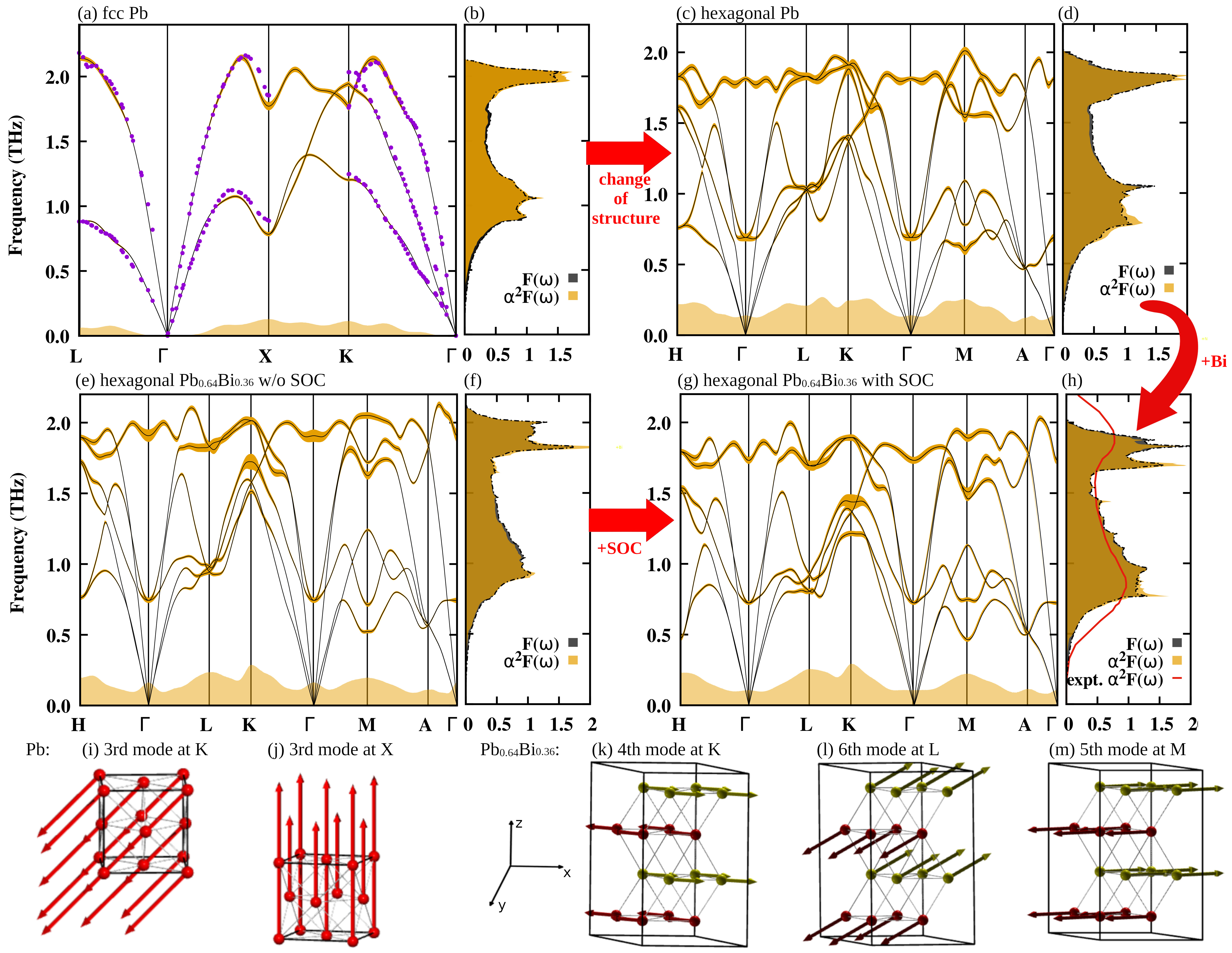}
\caption{The phonon and electron-phonon properties of $fcc$ Pb (a-b), hexagonal Pb (c-d), Pb$_{0.64}$Bi$_{0.36}$ alloy calculated  without SOC (e-f) and with SOC (g-h). The phonon dispersion relations in (a, c, e, g) are decorated with the phonon linewidth in the form of a fatband. Below the dispersion curves, the phonon linewidths summed over the bands are presented with a filled curve. The DOS phonon and Eliashberg functions are shown in (b, d, f, h). 
In case of Pb, the experimental data \cite{pb-ph} on the phonon dispersion relations are also presented with dots.
In panel (h), the Eliashberg function obtained from the tunneling experiment for Pb$_{0.65}$Bi$_{0.35}$ \cite{pbbi-history5} is plotted using the red line. 
Furthermore, the atomic displacements associated with strongly coupled phonon modes are schematically marked with arrows in panels (i-j) for Pb and (k-m) for Pb$_{0.64}$Bi$_{0.36}$.}\label{fig:elph}
\end{figure*}

In the next step, the Eliashberg function $\alpha^2F(\omega)$ is calculated as a sum of $\gamma_{{\bf q}\nu}$ over all phonon modes, weighted by the inverse of their frequency:
\begin{equation}
\label{eq:a2f}
\alpha^2F(\omega) = {1\over 2\pi N(E_F)}\sum_{{\bf q}\nu} 
                    \delta(\omega-\omega_{{\bf q}\nu})
                    {\gamma_{{\bf q}\nu}\over\hbar\omega_{{\bf q}\nu}}.
\end{equation}

The logarithmic average frequency, used in the Allen-Dynes formula, is defined as
\begin{equation}\label{eq:omlog2}
\omega_{\rm log}^{\alpha^2F}= \exp\left(\int_0^{\omega_{\mathsf{max}}} \alpha^2F(\omega) \ln\omega\frac{d\omega}{{\omega}} \left/ \int_0^{\omega_{\mathsf{max}}} 
{\alpha^2F(\omega)}\frac{d\omega}{{\omega}} \right. \right).
\end{equation}

Finally, the EPC constant $\lambda$ is calculated as the integral of the Eliashberg function divided by frequency
\begin{equation}\label{eq:lam2}
\lambda=2\int_0^{\omega_{\rm max}} \frac{\alpha^2F(\omega)}{\omega} \text{d}\omega.
\end{equation}  

For analysis purpose it is useful to calculate the integral $I$:
\begin{equation}
I=\int_0^{\omega_{\rm max}} \omega\cdot\alpha^2F(\omega){\rm d}\omega.
\label{eq:I}
\end{equation}
which is a frequency-independent measure of  electronic contribution to $\lambda$~\cite{gutowska2021,kuderowicz2022}.
This quantity does not depend directly on phonon frequency, as multiplication over $\omega$ cancels the $\alpha^2F(\omega) \propto 1/\omega$ dependence, and $I$ is proportional to the sum of phonon linewidths over the Brillouin zone (this parameter is closely related to McMillan-Hopfield parameter in monoatomic materials~\cite{mcmillan,hopfield}, see Supplemental Material~\cite{supplemental}). 
With the help of $I$ and the "average square" phonon frequency $\hat\omega^2$  defined in Eq.(\ref{eq:omega2}),
EPC constant may be expressed in an intuitive form as $\lambda = \frac{2I}{\hat\omega^2}$, becoming a ratio of contributions mainly electronic ($I$) and phononic (${\hat\omega^2}$) to $\lambda$.

The phonon dispersion relations $\omega({\bf q})$ and the phonon density of states $F(\omega)$ of $fcc$ Pb are shown in Fig. \ref{fig:elph}(a-b).
Its phonon spectrum consists of the three acoustic branches, which span the frequency range from 0 to 2.2~THz. The average frequency is 1.39 THz, as given in Table \ref{tab:freq}.
The calculated phonon dispersion relations are in good agreement with the experimental data, marked with dots in Fig. \ref{fig:elph}(a). This also includes the Kohn anomaly, seen as a deflection in the lowest band in the $\Gamma-K$ direction in $\bf q$-space, discussed in detail in Refs. \cite{pb-ph,pb-ph2}.
The phonon DOS in Fig.~\ref{fig:elph}(b) has a characteristic two-peak structure, with the first maximum around 1.0 THz contributed by the lowest transverse mode and the second, near 2.0 THz, by the longitudinal mode. For further considerations, we define high- and low-frequency modes as the parts of the spectrum that form the higher and lower peak of the DOS, respectively, and we set the border between them as $0.8\omega_{\rm max}$.

The strength of the electron-phonon interaction for a given phonon is visualized in Fig.~\ref{fig:elph}(a) with the help of the phonon linewidths. 
The interaction is fairly isotropic in the $\bf q$-space
and is the strongest for the highest (longitudinal) mode, especially in two $\mathbf{q}$-points, X and K.
The atomic vibrations, associated with these modes, are shown in real space in Fig. \ref{fig:elph} (i-j), which confirms their longitudinal nature. Stronger interaction for the highest phonon branch results in the Eliashberg function $\alpha^2F(\omega)$ slightly enhanced over the phonon DOS in panel (b).
The electron-phonon coupling constant $\lambda$ is equal to 1.44 (see Table~\ref{tab:tc}) and in 75\% it is contributed by the low-frequency modes.

Now, let us see the impact of the cubic-to-hexagonal structural transition on the phonons and the electron-phonon coupling. 
As in the hexagonal (nearly $hcp$) structure there are two atoms in the primitive cell, the number of phonon modes at each wavevector is doubled to six, making the phonon dispersion picture more complicated, with the interpenetrating acoustic and optical modes.
However, the essential effect is that if the Pb atoms were arranged to form the hexagonal structure of Pb$_{0.64}$Bi$_{0.36}$, the phonon frequencies would decrease [Fig. \ref{fig:elph}(c-d)]. The maximum frequency drops to 2.0 THz and the average to 1.26 THz. The general shape of the phonon DOS is kept.

\begin{table}[t]
\caption{The phonon frequency moments obtained with help of Eq.(\ref{eq:mom})--(\ref{eq:omlog2}).}
\label{tab:freq}
\begin{center}
\begin{ruledtabular}
\begin{tabular}{ l c c c c  }
	&$\langle\omega^1\rangle$ 	&	$\sqrt{\langle\omega^2\rangle}$	&	$\langle\omega\rangle$	&$\langle\omega_{\rm log}\rangle$	\\
	\hline
fcc Pb w/o SOC	&	1.35	&	1.44	&	1.53	&	1.24	\\
fcc Pb	&	1.21	&	1.29	&	1.39	&	1.11	\\
hexagonal Pb	&	1.07	&	1.16	&	1.26	&	0.97	\\
Pb$_{0.64}$Bi$_{0.36}$ w/o SOC &		1.22	&	1.3	&	1.38	&	1.13	\\
Pb$_{0.64}$Bi$_{0.36}$	&	1.11	&	1.18	&	1.26	&	1.02	\\
\end{tabular}
\end{ruledtabular}
\end{center}
\end{table}

To understand why the frequencies are lowered, we have to look at the force constants in these two variants of Pb. This is analyzed in more details in Supplemental Material~\cite{supplemental} where we show that the in-plane atom-atom force constants of the hexagonal structure are increasing but the coupling between nearest atoms in different planes is weaker, leading to a smaller total restoring force when the atom is displaced.  This leads to lower phonon frequencies.

The phonon linewidths in the hexagonal phase become larger at higher frequencies, and the "electronic" parameter $I$ slightly grows, from 1.32~THz$^2$ in the $fcc$ Pb to  1.37 THz$^2$ in the hexagonal structure, see Table~\ref{tab:tc}. 
However, when it comes to the Eliashberg function, in which the phonon linewidth is divided by frequency, it also deviates from phonon DOS at low frequencies (around 0.7 THz). It indicates an enhanced electron-phonon coupling in this regime, coming mainly from two low-frequency optical modes at $\Gamma$. 
This results in a decrease of $\hat\omega^2$ from 1.79 in $fcc$ to 1.35 THz$^2$ in the hexagonal phase, leading to a significant increase in the electron-phonon coupling constant: from 1.47 to 2.03.
The contribution of the low-frequency part of the phonon spectrum is now as large as 92\%. 
This shows the primary importance of the cubic-to-hexagonal transition in shaping the phonons and the strong electron-phonon coupling in the system. 

{The electron doping effect itself, inherent to the transformation of the cubic Pb structure to hexagonal Pb$_{0.64}$Bi$_{0.36}$, further modify the phonon structure and electron-phonon coupling, but to a lesser degree. 
Figures \ref{fig:elph}(g-h) show the phonon spectrum and the Eliashberg function for Pb$_{0.64}$Bi$_{0.36}$. Additionally, the Eliashberg function extracted from the tunneling data for an alloy of very similar concentration, Pb$_{0.65}$Bi$_{0.35}$ \cite{pbbi-history5}, is shown. It agrees with the calculated one, validating the chosen calculation method.}
{The electron-phonon coupling constant {of Pb$_{0.64}$Bi$_{0.36}$ alloy} is equal to $\lambda = 2.05$ and is contributed by low-frequency ($< 1.65$ THz) modes in 84\%. With respect to $fcc$ Pb, this 40\% growth of $\lambda$ from 1.47 results both from the increase in the electronic contribution to $\lambda$, measured by the parameter $I$ (which grows in 10\%, see Table~\ref{tab:tc}), and from the decrease in the phonon frequencies ($\hat\omega^2$ decreases in more than 20\%). 
However, the electron-phonon properties of the alloy are rather similar to those of hypothetical Pb in hexagonal structure. Despite the heavier mass of Bi than Pb, the average frequency of Pb$_{0.64}$Bi$_{0.36}$ alloy is the same (1.26 THz) and other frequency moments are even slightly larger (see Tables~\ref{tab:freq} and \ref{tab:tc}, $\hat\omega^2$ increase to 1.41 THz$^2$). 
The additional 0.36 electrons provided by Bi slightly stiffens the atom-atom bonds, which compensates for the mass increase effect. 
It is seen in the value of force constants~\cite{supplemental}, which are slightly larger in case of Pb$_{0.64}$Bi$_{0.36}$ alloy and increase some of the phonon frequencies. 
However, as the electronic contribution $I$ also increase in Pb$_{0.64}$Bi$_{0.36}$ to 1.45~THz$^2$, this compensates for the slight increase in phonon frequencies leading to a high value of $\lambda = 2.05$. Thus, in Pb$_{0.64}$Bi$_{0.36}$ alloy, both the Bi electron doping effect and transition to hexagonal structure cooperate in enhancing the electron-phonon coupling parameter $\lambda$ over the one observed in metallic Pb.
}

{Looking now at the anisotropy of electron-phonon interaction in the studied systems, one of the possible indications of anisotropy is a deviation of the shape of Eliashberg function from the shape of the phonon DOS, meaning that the strength of the electron-phonon coupling is mode-dependent.}  As in the case of both hexagonal and $fcc$ Pb, $\alpha^2F$ {of  Pb$_{0.64}$Bi$_{0.36}$ alloy} deviates from the shape of the phonon DOS at high frequencies (around 1.7 THz) due to large phonon linewidths of the highest mode at the $L$ $\bf q$-point. The second enhancement is observed at low frequency, as in hexagonal Pb. It comes mainly from the phonons at the $\Gamma$ point. However, in Pb$_{0.64}$Bi$_{0.36}$ additional large phonon linewidths appear at the $K$ and $M$ $\mathbf{q}$-points. 
The modes with mentioned $\mathbf{q}$ vectors are shown in real space in Fig. \ref{fig:elph}(k-m). All of them are longitudinal, and they stretch the bonds between different atomic planes.

 \begin{table}[t]
\caption{The electron-phonon and superconducting properties of Pb and Pb$_{0.64}$Bi$_{0.36}$ alloy in terms of EPC parameter $\lambda$, integral $I$ (THz) defined with Eq. \ref{eq:I}, two characteristic Eliashberg function moments: $ \hat\omega^2$ (THz$^2$, Eq. \ref{eq:omega2}) and $\omega_{\rm log}^{\alpha^2F}$ (THz, Eq. \ref{eq:omlog2}), critical temperature $T_c$ (K, Eq. \ref{eq:tc-modified}).}
\label{tab:tc}
\begin{center}
\begin{ruledtabular}
\begin{tabular}{ l c c c c  c c}
	&		$\lambda$	&	$I$	&	$\hat\omega^2$	&	$\omega_{\rm log}^{\alpha^2F}$	&	T$_c$	\\
\hline											
fcc Pb w/o SOC	&	1	&	1.17	&	2.43	&	1.38	&	4.86	\\
fcc Pb	&	1.47	&	1.32	&	1.79	&	1.19	&	7.1	\\
hexagonal Pb	&	2.03	&	1.37	&	1.23	&	0.99	&	8.21	\\
Pb$_{0.64}$Bi$_{0.36}$ w/o SOC	&	1.56	&	1.37	&	1.72	&	1.18	&	7.48	\\
Pb$_{0.64}$Bi$_{0.36}$	&	2.05	&	1.45	&	1.43	&	1.06	&	8.67	\\
\end{tabular}
\end{ruledtabular}
\end{center}
\end{table}

Finally, we analyze the spin-orbit coupling effect on phonons and the electron-phonon interaction.
The phonons calculated without SOC are shown in Fig. \ref{fig:elph}(e-f) and shall be compared with those calculated with SOC in Fig. \ref{fig:elph}(g-h). 
Although the shape of DOS is similar in both these cases, almost all phonon modes have lower frequencies when spin-orbit coupling is included. This is especially seen for the modes with large phonon linewidths. 
The average frequency is reduced from 1.38 THz without SOC to 1.26 THz with SOC, see Table~\ref{tab:freq}. 
A similar effect of SOC is observed in $fcc$ Pb where the average frequency is reduced from 1.53 THZ without SOC to 1.39 THz with SOC\footnote{Compare Fig. \ref{fig:elph}(a-b) with Fig. S3 in the Supplemental Material~\cite{supplemental}, which shows the phonon and electron-phonon properties of Pb calculated without SOC.}.
The reason for this behavior is the above-mentioned SOC-induced contraction of the $p$-orbital wavefunction, which reduces the electron density between the atoms (see Fig. \ref{fig:charge}) and weakens the bonds (see force constants in the Supplemental Material~\cite{supplemental}).
This effect is also accompanied by a smaller increase in the electronic contribution to $\lambda$, as shown in Table~\ref{tab:tc}, for both, $fcc$ Pb and Pb-Bi alloy. 
As a result, the electron-phonon coupling constant $\lambda$ is increased by spin-orbit coupling in 47\% in Pb and 31\% in Pb$_{0.64}$Bi$_{0.36}$. The value of $\lambda$ obtained with SOC agrees well with the value calculated as a renormalization factor of the electronic specific heat, as seen when comparing the {relativistic values of $\lambda$ of Pb and Pb$_{0.64}$Bi$_{0.36}$ in Table~\ref{tab:tc} to Table~\ref{tab:gammy}.}

\subsection{Superconductivity}

In the simplified McMillan-Allen-Dynes approach superconducting critical temperature [Eq. (\ref{eq:tc-modified})] depends on three parameters:
the electron-phonon coupling constant $\lambda$, average logarithmic frequency $\omega_{\rm log}^{\alpha^2F}$ and the retarded Coulomb pseudopotential parameter $\mu^*$.

As discussed above, $\lambda$ evolves from 1.47 for $fcc$ Pb, through 2.03 for hexagonal Pb to 2.05 for Pb$_{0.64}$Bi$_{0.36}$, while the logarithmic frequency changes from 1.19 THz, through 0.99 THz to 1.06 THz, respectively (see Table~\ref{tab:tc}).
This shows that the key to the record strong coupling superconductivity of Pb$_{0.64}$Bi$_{0.36}$ alloy is its transition to the hexagonal structure. The electron doping effect is also beneficial; not only because it triggers the $fcc$-hexagonal transition but additionally increases $\omega_{\rm log}^{\alpha^2F}$ in the hexagonal phase, while maintaining the large value of $\lambda$.

Within the Allen-Dynes formula, taking $\mu^* = 0.10$ the calculated critical temperatures are 7.1 K, 8.21 K and 8.67 K for $fcc$ Pb, hexagonal Pb and Pb$_{0.64}$Bi$_{0.36}$, in remarkable agreement with the experimental data (7.2~K for $fcc$ Pb and 8.6 K for Pb$_{0.64}$Bi$_{0.36}$) 
\footnote{If $f_1$ and $f_2$ strong-coupling corrections in Eq. (\ref{eq:tc-modified}) are neglected, the critical temperature is underestimated: 6.41 K for $fcc$ Pb, through 6.93 K for hexagonal Pb to 7.44 K for Pb$_{0.64}$Bi$_{0.36}$.}.
All these results are shown in Table \ref{tab:tc}. 
Although the value of $T_c$ is well reproduced by the Allen-Dynes formula, the superconductivity in these systems is not a simple single-gap s-wave like, as we show in the next part of this work, where we determine the superconducting properties of Pb$_{0.64}$Bi$_{0.36}$ within the isotropic Eliashberg formalism~\cite{Eliashberg1960}, as well as by using the density functional theory for superconductors (SCDFT). 
The $fcc$ Pb is also discussed as a reference material.

\subsubsection{Eliashberg formalism}

The isotropic Eliashberg equations, defined in the imaginary axis, are given by~\cite{Eliashberg1960}
\begin{eqnarray}
Z(i\omega _n)&=&1+\frac{\pi k_B T}{\omega _n} \sum _{n'} \frac{\omega _{n'}}{R(i\omega _{n'})} K(i(\omega_{n}-\omega_{n'})) \nonumber \\
 Z(i\omega _n) \Delta(i\omega _n) &=& \pi k_B T \sum _{n'} \frac{\Delta(i\omega _{n'})}{R(i\omega _{n'})} \times  \label{eq:eliash} \\
 &&[ K(i(\omega_n-\omega_{n'}))-\mu^* \theta ( \omega _c - |\omega _{n'}| )],\nonumber
\end{eqnarray}
where $Z(i\omega_n)$ is the mass renormalization function, $\Delta(i\omega _n)$ is the superconducting order parameter, $i\omega_n=i(2n+1)\pi k_B T$ 
are fermionic Matsubara frequencies where $n\in \mathbb{Z}$, $\theta(\omega)$ is the Heviside function, $k_B$ is the Boltzmann constant, $T$ is  
temperature and 
\begin{equation}
R(i\omega _n) = \sqrt{\omega _n ^2 + \Delta^2(i\omega _n)}. 
\end{equation}
The kernel of the electron-phonon interaction takes the form
\begin{equation}
 K(i(\omega_n-\omega_{n'}))=\int _0^{\infty} d \omega \frac{2 \omega \alpha ^2 F(\omega)}{(\omega _n - \omega _{n'})^2+\omega ^2},
\end{equation}
where $\alpha ^2 F(\omega)$ is the isotropic Eliashberg spectral function.
Equations (\ref{eq:eliash}) are solved with a method described in Refs.~\cite{kuderowicz2022,kuderowicz2021} in self-consistent manner by the use of the following values of the parameters: number of the Matsubara frequencies $M=8000$ and cut-off energy $\omega_c=8\omega_{max}$, where $\omega_{max}$ is the upper limit of the phonon frequency.

The pseudopotential parameter $\mu^*$ is used to account for the effects of the retarded electron-electron depairing interactions. 
Although calculations of $\mu^*$ from the {\it ab-initio} methods are possible, they require more sophisticated numerical methods~\cite{pickett,mgb2-mu}. For this reason, it is often treated as a parameter whose value is determined based on the requirement that the solution gives the critical temperature in agreement with the experimental $T_c$. 
This procedure enables us to discuss the thermodynamic properties of the material as a function of $T/T_c$ and to compare them with the experimental data. 
Fully ab-initio calculations of $T_c$ from the SCDFT method, without assumptions on $\mu^*$, are discussed in the next paragraph.

\begin{figure}[t]
\includegraphics[width=.45\textwidth]{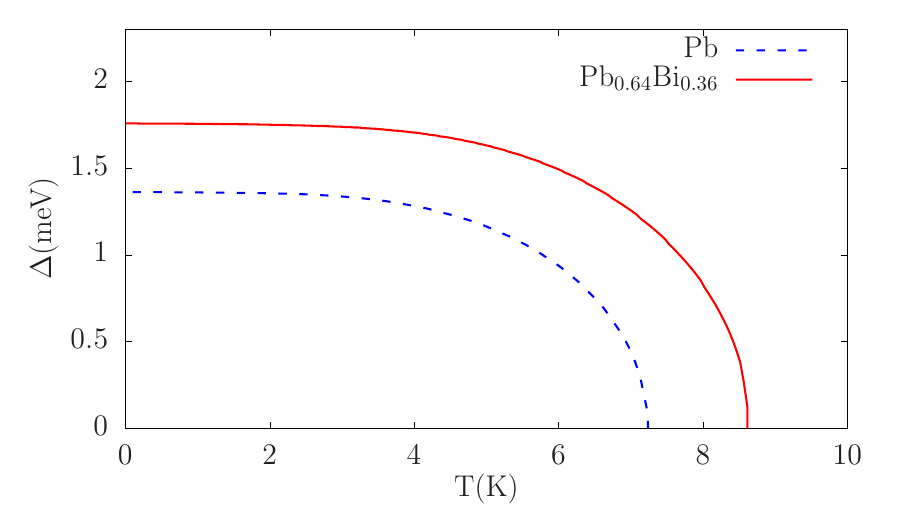}
\caption{Temperature dependence of superconducting gap for Pb and Pb$_{0.64}$Bi$_{0.36}$ calculated within the isotropic Eliashberg approach.}\label{fig:gap}
\end{figure}

The evolution of the superconducting gap $\Delta(n=0)$  with temperature for Pb and Pb$_{0.64}$Bi$_{0.36}$ is presented in Fig. \ref{fig:gap}. 
To obtain the experimental $T_c$ of Pb ($T_c=7.2$~K) and Pb$_{0.64}$Bi$_{0.36}$ ($T_c=8.6$~K) the Coulomb pseudopotentials of $\mu^*=0.117$ and $0.134$ have to be taken. 
Note that the $\mu^*$ parameter used here is not exactly the same as used in the Allen-Dynes equation (\ref{eq:tc-modified}) due to the dependence of the solution of Eq.(\ref{eq:eliash}) on the cutoff frequency, which is discussed below.

In Fig.~\ref{fig:gap} we can see that the superconducting gap $\Delta(T=0\,{\rm K})$ of Pb is equal to $1.36$~ meV, which gives 
$\frac{2\Delta(0)}{k_BT_c}=4.39$, 
higher than the BCS value $\frac{2\Delta_{\rm BCS}}{k_BT_c}=3.54$. 
This result is in very good agreement with the value obtained from the tunneling data \cite{pbbi-history5} (1.36 meV, see Table \ref{tab:sctk}). 
In case of Pb$_{0.64}$Bi$_{0.36}$ alloy the superconducting gap at $T=0$ has been evaluated to $1.76$ meV, which gives $\frac{2\Delta(0)}{k_BT_c}=4.76$ also much higher than the BCS limit and slightly smaller than $4.98$ determined from the tunneling data \cite{allen-dynes} for Pb$_{0.65}$Bi$_{0.35}$. 

In the framework of Eliashberg model the difference in the electronic specific heat determined in the superconducting and normal state $\Delta C_e=C_e^S-C_e^N$ can be expressed as
\begin{equation}
 \frac{\Delta C_e(T)}{k_B N(E_F)} = -\frac{1}{\beta}\frac{d^2 (\frac{\Delta F}{N(E_F)})}{d(k_BT)^2},
\label{eq:Cv}
\end{equation}
with the specific heat in the normal state given by
\begin{equation}
\frac{C_e^N(T)}{k_B N(E_F)}=\frac{\pi^2}{3} k_B T (1+\lambda),
\end{equation}
where $\lambda$ is the electron-phonon coupling constant and $N(E_F)$ corresponds to the density of states at the Fermi level.
In Eq.(\ref{eq:Cv}), $\Delta F$ is the free energy difference between the superconducting and normal state,
 \begin{eqnarray}
  \frac{\Delta F}{N(E_F)}&=&-\pi k_BT \sum _{n} \left ( \sqrt{\omega_n^2+\Delta _n^2} - |\omega _n|\right ) \times \nonumber \\
  && \left ( Z^S(i\omega _n) - Z^N(i\omega _n) \frac{|\omega _n|}{\sqrt{\omega _n^2+\Delta_n^2}} \right ),
 \end{eqnarray}
 where $Z^S$ and $Z^N$ denote the mass renormalization factors for the superconducting (S) and normal (N) states, respectively. 
\begin{figure}[t]
\includegraphics[width=.49\textwidth]{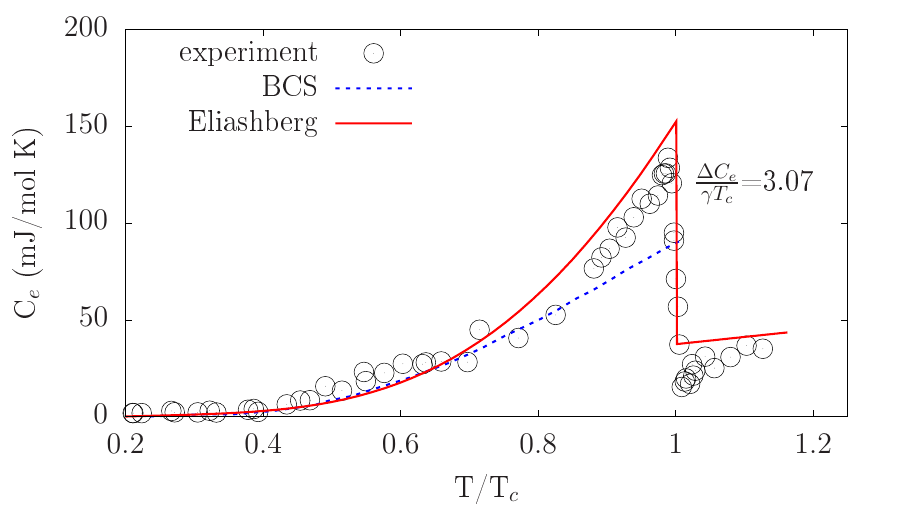}
\caption{Electronic specific heat $C_e$ as a function of temperature $T$ determined from the BCS theory (blue dashed line) and the Eliashberg equations (red dashed line) for Pb$_{0.64}$Bi$_{0.36}$ alloy. The experimental data are marked by black dots. }
\label{fig:cvel}
\end{figure}

Figure~\ref{fig:cvel} presents the temperature dependence of the specific heat with the experimental data plotted by dots. For comparison, the BCS result, which predicts the exponential behavior of the electronic specific heat at low temperatures, is also displayed by the blue line. 
As expected, Pb$_{0.64}$Bi$_{0.36}$ alloy which can be classified as a strong-coupled superconductor, exhibits non-BCS behavior of $C_e$ with the experimental value of the specific heat jump $\Delta C_e/\gamma T_c=2.91$ much higher than the BCS value $1.43$. Importantly, the inclusion of the retardation within the Eliashberg model allows reproducing $\Delta C_e/\gamma T_c$ at the level of $3.07$, close but above the experimental value. 
However, note that the $C_e(T)$ curve predicted by the isotropic formalism differs from the experimental points. 
In particular, it does not reproduce the change in curvature that appears around $0.8T_c$ (7~K). This type of behavior strongly suggests the multiband character and anisotropy of the superconducting gap in this compound.

\begin{figure}[t]
\includegraphics[width=.47\textwidth]{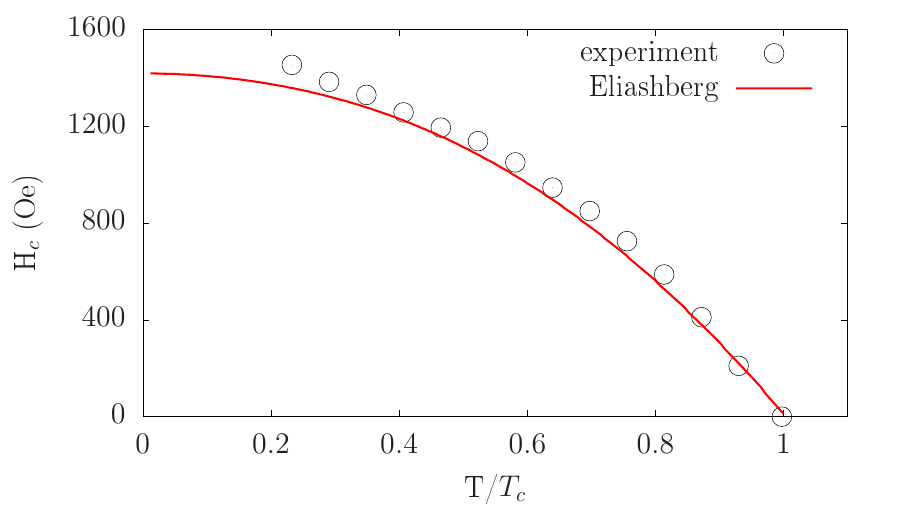}
\caption{Thermodynamic critical field $H_{c}(T)$ as a function of normalized temperature $T/T_c$ calculated from the Eliashberg model. Dots represent experimental data. }
\label{fig:Hc}
\end{figure}

Similar behavior
can also be seen in the temperature dependence of the thermodynamic critical field, presented in Fig.~\ref{fig:Hc}. Here, the critical field within the Eliashberg model has been determined from the formula $\frac{H_c}{\sqrt{N(E_F)}}=\sqrt{-8\pi \frac{\Delta F}{N(E_F)}}$. 

At low temperatures, the experimental dependence $H_{c}(T)$ exhibits an unusual increase which is not predicted by the Eliashberg theory. Interestingly, also in the mid-temperature range, the model does not reproduce the curvature of $H_{c}(T)$ obtained in the experiment and the deviation is observed below $0.8T_c$.
A similar situation is observed for
the upper critical field $H_{c2}$, which we calculated within the isotropic Eliashberg approach ~\cite{Carbotte1990,kuderowicz2021} - relevant formulas are given in the Supplemental Material~\cite{supplemental}.

\begin{figure}[t]
\includegraphics[width=.47\textwidth]{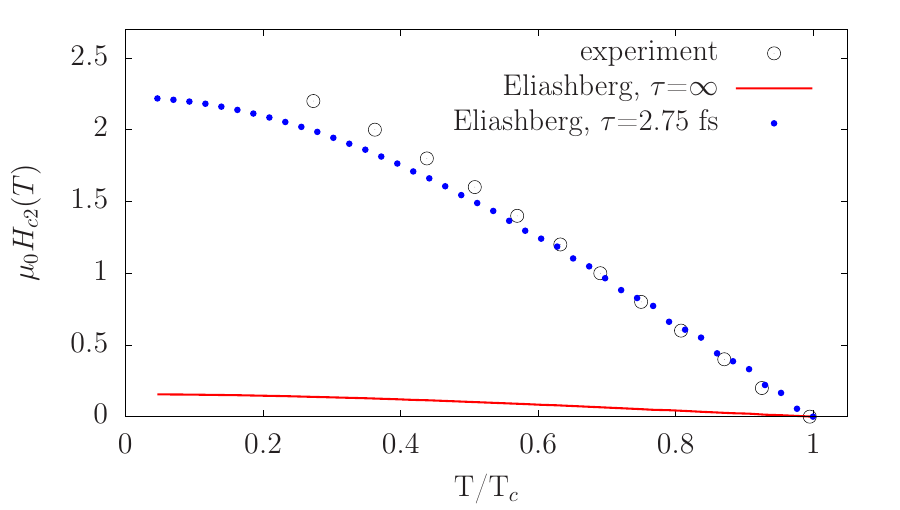}
\caption{Temperature dependence of the upper critical field $H_{c2}(T)$  calculated within the Eliashberg model with $\tau = 2.75\times 10^{-15}$ s. The red line shows the result for the clean limit (no scattering, $\tau = \infty$). Dots represent experimental data. }
\label{fig:Hc2}
\end{figure}

In the calculations of $H_{c2}$, for the additional required input parameter, which is the electronic scattering time~\cite{supplemental}, the value of $\tau = 2.75$~fs, determined in
Sect.~\ref{sec:disorder} was used.
This value of $\tau$ places the material near the dirty limit (the clean limit, in which electron scattering is neglected, corresponds to $\tau = \infty$). 
The calculated temperature dependence of $H_{c2}(T)$ is shown in Fig.~\ref{fig:Hc2} with the experimental data marked by dots. For comparison, results evaluated for the clean limit has also been displayed. 
As seen, the measured critical field is above $2$~T at low temperatures, and results from the alloy scattering effects, as $H_{c2}(T=0)$ in the clean limit is estimated to be below $0.2$~T. 
Our calculations for the $\tau = 2.75$ fs scattering time
reproduce the experimental curve quite well for $T/T_c > 0.7$.  
However, in the low temperature range, the theoretical curve $H_{c2}(T)$ deviates from the experimental points, which is another indication that the single-gap s-wave model is not appropriate to capture all features of the system considered.

In summary, the isotropic Eliashberg formalism reproduces quite well the strong-coupling experimental results for the characteristic ratios of $2\Delta/k_BT_c$ or $\Delta C_e/\gamma T_c$, however, 
the temperature dependence of the specific heat or critical field deviate from the experiment.
This strongly indicates the multiband character of superconductivity in the Pb-Bi alloy, probably accompanying the anisotropy of the energy gap.

\subsubsection{$T_c$ from SCDFT}

\begin{figure}[t]
    \centering
\includegraphics[width=0.47\textwidth]{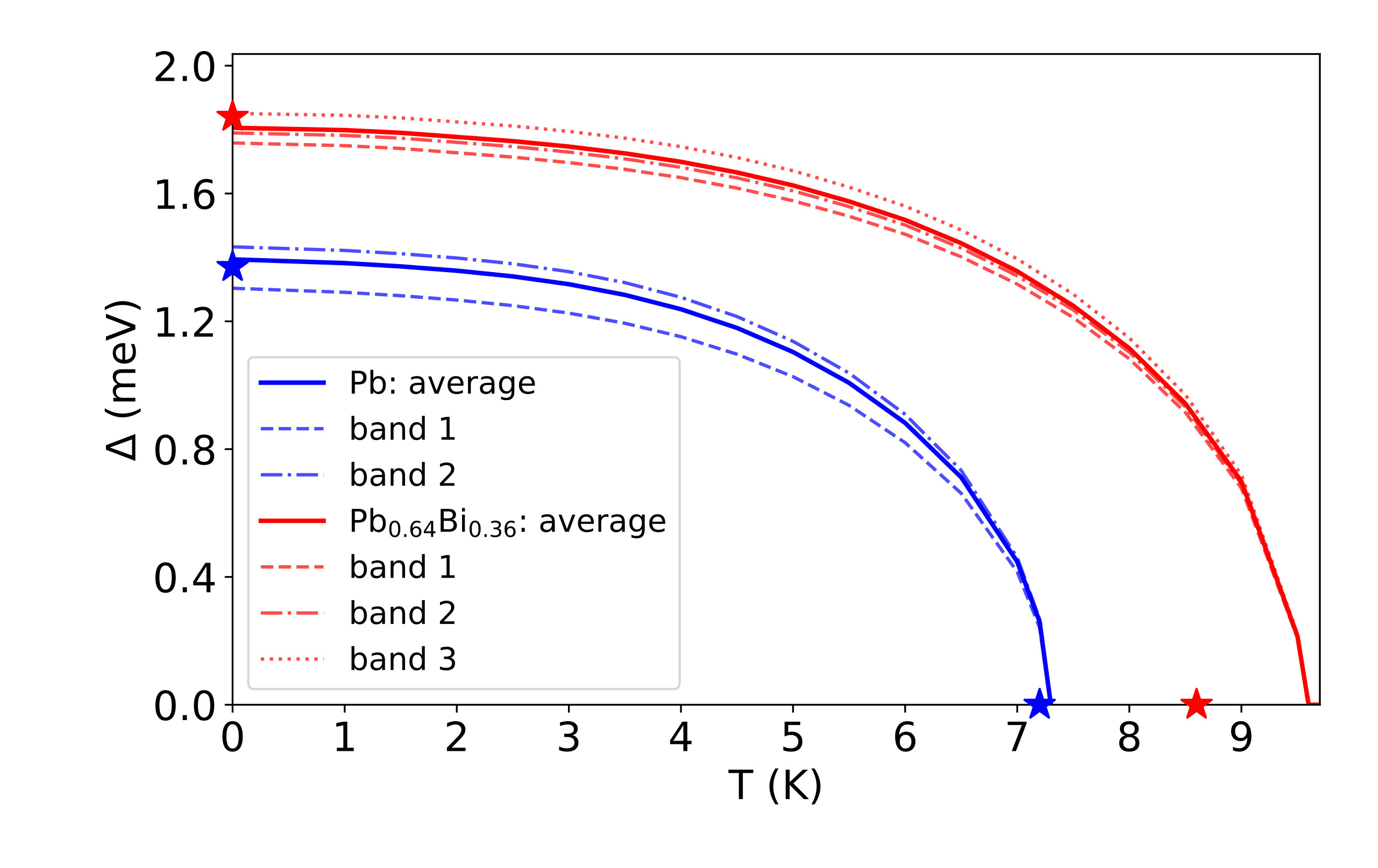}
    \caption{Temperature dependence of superconducting gaps at each Fermi surface sheet (dashed lines) and averaged over the Fermi surface (solid lines) of Pb (blue color) and Pb$_{0.64}$Bi$_{0.36}$ (red color). The experimental values of $\Delta(T=0)$ and $T_c$ are marked with stars.}
    \label{fig:gap_bands}
\end{figure}

The density functional theory for superconductors (SCDFT)~\cite{Oliveira1988,Luders2005,Kawamura2017,sctk} as implemented in the {\sc Superconducting Toolkit} (SCTK) package \cite{sctk} was applied to investigate the structure of the superconducting gap in the wave vector space, as well as to calculate $T_c$ taking into account the effects of repulsive Coulomb interactions without the need to make assumptions about the value of $\mu^*$.
This approach is based on the solution of the following self-consistent equations~\cite{Kawamura2017} for the superconducting gap function $\Delta_{n\mathbf{k}}$:
\begin{align}        \label{eq:gap3}
    \Delta_{n\mathbf{k}} = & -\frac{1}{2}\sum_{n'\mathbf{k}'} \frac{K_{n\mathbf{k}n'\mathbf{k}'}(\xi_{n\mathbf{k}},\xi_{n'\mathbf{k}'})}{1+Z_{n\mathbf{k}}(\xi_{n\mathbf{k}})} \nonumber \\
    & \times \frac{\Delta_{n'\mathbf{k}'}}{\sqrt{\xi_{n'\mathbf{k}'}^2+\Delta_{n'\mathbf{k}'}^2}} \tanh \frac{\sqrt{\xi_{n'\mathbf{k}'}^2+\Delta_{n'\mathbf{k}'}^2}}{2T},
\end{align}
where $\xi_{n\mathbf{k}}$ is the Kohn-Sham eigenvalue in the band $n$ at the wavevector $\mathbf{k}$ (at $E_F=0$), $K\equiv K^{ee}+K^{ep}$
consists of electron-electron ($K^{ee}$) and electron-phonon interaction kernels ($K^{ep}$), and $Z$ is the renormalization function. The screened Coulomb interaction is calculated with the random phase approximation (RPA), for more details, see Refs.~\cite{Kawamura2017, sctk}.

\begin{figure*}[ht] 
\includegraphics[width=\textwidth]{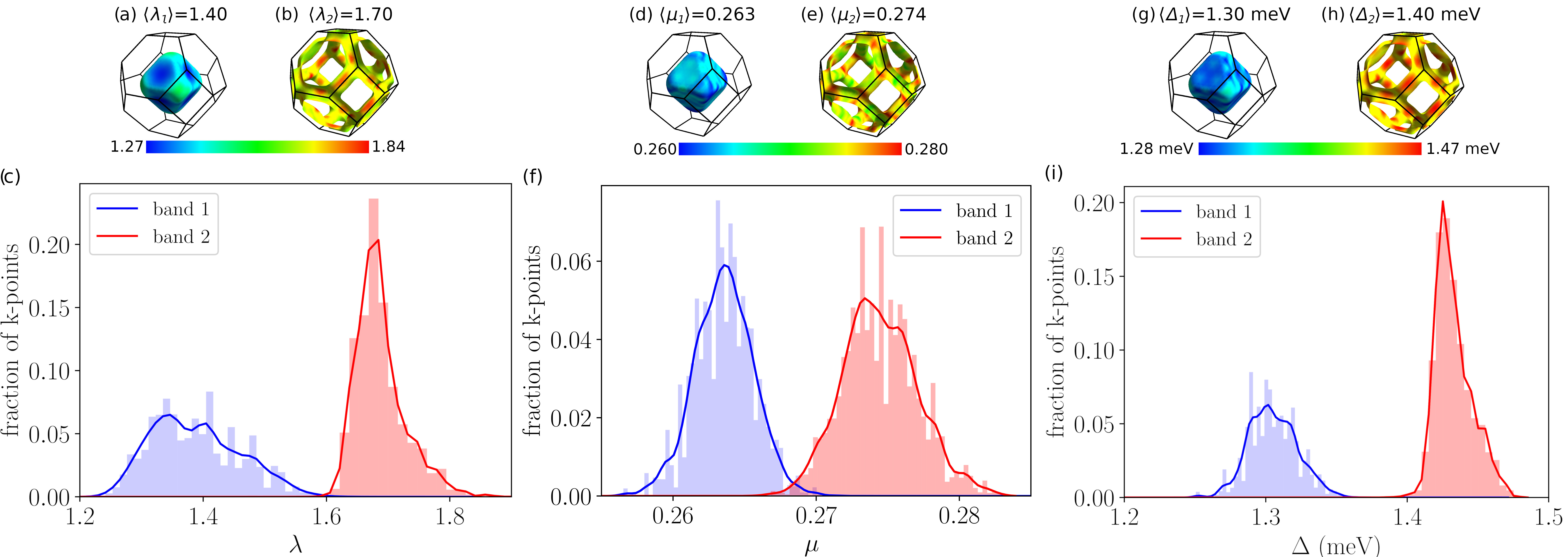}
\caption{Pb: the $n\mathbf{k}$-dependent (a-b) EPC parameter $\lambda_{nk}$; (d-e) Coulomb screened potential $\mu_{nk}$; (g-h) superconducting gap $\Delta_{nk}$ (meV) plotted at the Fermi surface. Histograms of these parameters are shown in panels (c, f, i).} \label{fig:aniso-pb}
\end{figure*}

\begin{figure*}[ht] 
\includegraphics[width=\textwidth]{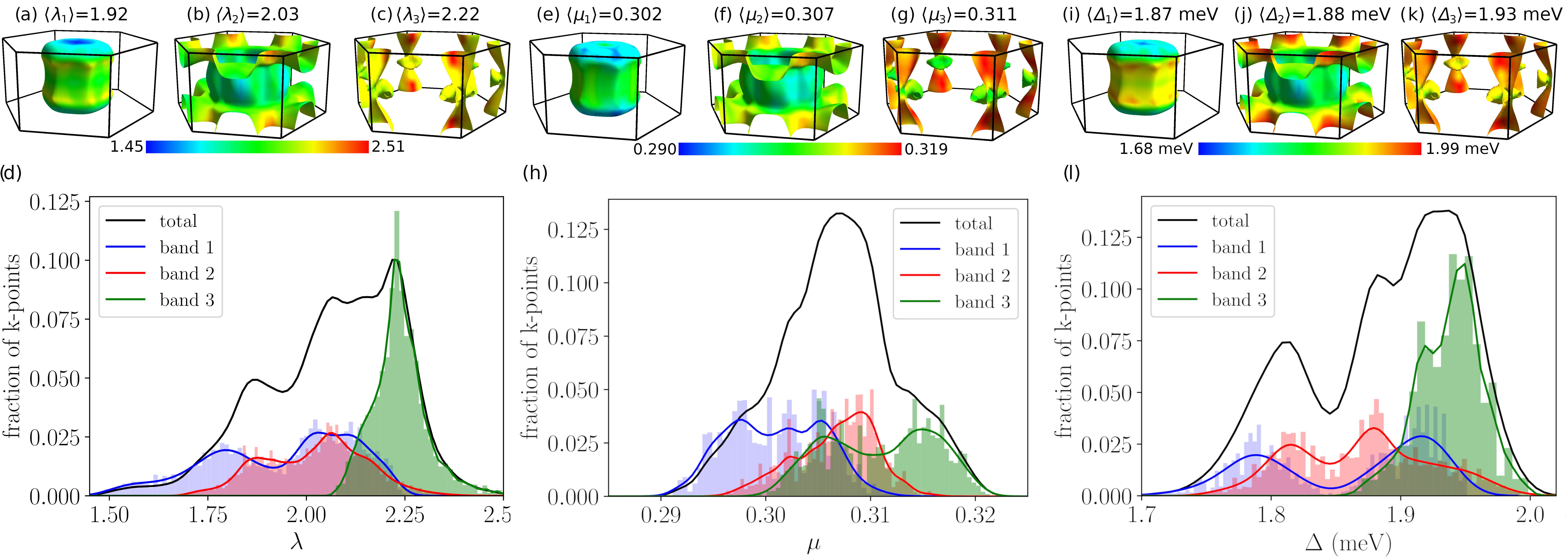}
\caption{Pb$_{0.64}$Bi$_{0.36}$: 
the $n\mathbf{k}$-dependent (a-c) EPC parameter $\lambda_{nk}$; (e-g) Coulomb screened potential $\mu_{nk}$; (i-k) superconducting gap $\Delta_{nk}$ (meV) plotted at the Fermi surface. Histograms of these parameters are shown in panels (d, h, l).} \label{fig:aniso-pbbi}
\end{figure*}

The critical temperature $T_c$ is calculated from the temperature dependence of the gap averaged over the $\mathbf{k}$ vector, shown in Fig. \ref{fig:gap_bands}.
The distribution of key quantities on the Fermi surface, i.e., the electron-phonon coupling parameter $\lambda_{n\mathbf{k}}$, the screened Coulomb interaction parameter $\mu_{n\mathbf{k}}$ and the superconducting gap $\Delta_{n\mathbf{k}}$ is shown in Fig.~\ref{fig:aniso-pb} for Pb and in Fig.~\ref{fig:aniso-pbbi} for the Pb-Bi alloy.
The screened Coulomb interaction parameter $\mu_{n\mathbf{k}}$ is obtained from the electron-electron interaction kernel $K^{ee}_{n\mathbf{k}n'\mathbf{k}'}$ [see Eq. (\ref{eq:gap3})] by summing over $\{n'\mathbf{k}'\}$ on the Fermi surface, 
and $\Delta_{n\mathbf{k}}$ is the solution of the gap equation.

The calculated average value of ${\Delta}(T=0)$ agrees very well with the experimental data for both Pb and Pb$_{0.64}$Bi$_{0.36}$, marked by stars on the vertical axis in Fig.~\ref{fig:gap_bands} and listed in Table \ref{tab:sctk}.
The theoretical critical temperatures are 7.3 K for Pb and 9.6 K for Pb$_{0.64}$Bi$_{0.36}$. 
Thus, in contradiction to Pb for which the theoretical $T_c$ is very close to the experimental one, the critical temperature for Pb$_{0.64}$Bi$_{0.36}$ is quite clerly overestimated, compared to the experimentally measured $8.6$ K. 
We argue in the following that to a large degree it is related to disorder-induced electron scattering, which reduces $T_c$ in the Pb-Bi alloy and is not taken into account in the SCDFT calculations, based on the VCA electronic structure. 
Both materials are confirmed to be strongly-coupled superconductors, with $2\Delta/k_BT_c = 4.42$ (Pb) and 4.54 (Pb-Bi) (when the SCDFT $T_c$ is used) and 5.04 (when the experimental $T_c$ is used).

\subsubsection{Anisotropy of superconducting gap} 
In recent years there has been a discussion on the structure of the superconducting gap in Pb
in both experimental and theoretical works. 
Anisotropic two gaps were reported in Refs.~\cite{pb-gap1,pb-gap2}. The two pieces of the Fermi surface of Pb were proposed to give rise to separate gaps in the ranges of 1.16-1.28 meV (small gap) and 1.37-1.43 meV (large gap) \cite{pb-gap2}. On the other hand, recently the isotropic nature of the gap has been proposed \cite{pb-gap} in the analysis of the critical field deduced from the $\mu$SR measurement.
In case of Pb$_{0.64}$Bi$_{0.36}$ non such analysis has been performed yet.

Returning to Figs. \ref{fig:aniso-pb} and \ref{fig:aniso-pbbi} with the $\mathbf{k}$- resolved
$\lambda_{n\mathbf{k}}$, $\mu_{n\mathbf{k}}$ and $\Delta_{n\mathbf{k}}$
we see that according to the SCDFT calculations the superconductivity in both materials has an anisotropic and multiband nature, which is a consequence of the large distribution of the values of the electron-phonon coupling parameter $\lambda_{n\mathbf{k}}$ on the Fermi surface. Moreover, in both cases the distribution of $\lambda_{n\mathbf{k}}$ and $\Delta_{n\mathbf{k}}$ on the Fermi surface follows the orbital $p$ character, shown in Figs. \ref{fig:elpb} and \ref{fig:elsoc}.

For Pb,  $\lambda_{n\mathbf{k}}$, presented in Fig.~\ref{fig:aniso-pb}(a-c) spans the range of 1.27 to 1.80 and has a bimodal distribution on the two FS sheets, with an average value of $\langle \lambda_1 \rangle = 1.40$ and {$\langle \lambda_2 \rangle = 1.70$}.
{These averages over the Fermi surface for given band $n$ are calculated as}\footnote{Average values of $\mu_{n\mathbf{k}}$ and $\Delta_{n\mathbf{k}}$ are computed in the same way.}
\begin{equation}
   \langle \lambda_n \rangle= \frac{\sum_{\bm k} \lambda_{n{\bm k}} \cdot  \delta(E_{n{\bm k}}-E_F) }{ \sum_{\bm k} \delta(E_{n{\bm k}}-E_F)} =  \frac{\sum_{\bm k} \lambda_{n{\bm k}} \cdot \delta(E_{n{\bm k}}-E_F) }{ N_n(E_F)},
\end{equation}
using the tetrahedron method. The global average value of electron-phonon coupling constant is computed as
\begin{equation}
\lambda=\frac{\sum_{{\bm k}n} \lambda_{n{\bm k}} \delta (E_{n{\bm k}}-E_F) }{ \sum_{{\bm k}n} \delta(E_{n{\bm k}}-E_F)} =\frac{\sum_n \lambda_n N_n(E_F)}{ N(E_F)},
\end{equation} 
and is equal to  1.59, in a good agreement with the value of 1.47 calculated from the isotropic Eliashberg function in Sec.~\ref{sect:elph}.

The screened Coulomb pseudopotential $\mu_{n\mathbf{k}}$ is presented in Fig.~\ref{fig:aniso-pb}(d-f). 
By integrating $\mu_{n\mathbf{k}}$ over FS we obtain the average screened Coulomb repulsion parameter for each sheet $\langle\mu_n\rangle$, shown in Table~\ref{tab:sctk}, which is a dimensionless product of the average interaction kernel times the density of states at $E_F$~\cite{sctk}.
It is generally smaller for the first FS sheet, $\langle{\mu_1}\rangle = 0.263$, compared to the second, where $\langle{\mu_2}\rangle = 0.274$. This is related to the lower DOS value and the larger $s$ and $p_{1/2}$ character of the first sheet, as shown in Fig.~\ref{fig:elpb}, and follows the $\lambda_{n\mathbf{k}}$ distribution. 
As a consequence, the superconducting gap $\Delta_{n\mathbf{k}}$  in Fig.~\ref{fig:aniso-pb}(g-i) is generally smaller on the first FS sheet and larger on the second FS sheet. 
Consequently, the histogram of $\Delta_{n\mathbf{k}}$ is bimodal, corresponding to two separated superconducting gaps, with an average value of 1.30 meV ($\frac{2\Delta_1}{k_BT_c}=4.19$) and the second with an average value of 1.44 meV ($\frac{2\Delta_2}{k_BT_c}=4.64$). Our results generally agree with the theoretical calculations of Ref. \cite{floris2007two}, with our $T_c$ being closer to the experimental result.

\begin{figure}[t]
\includegraphics[width=.50\textwidth]{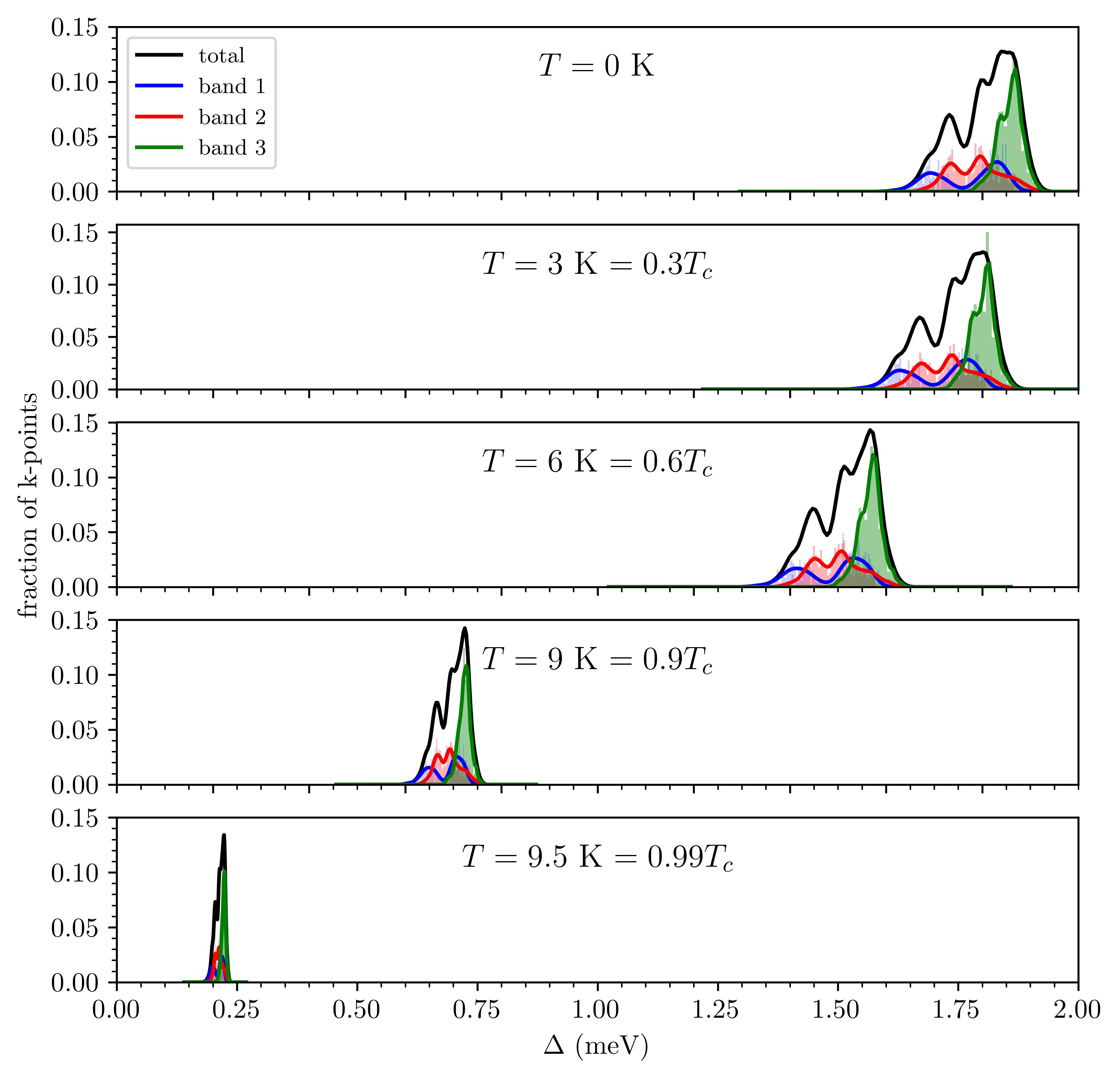}
\caption{The temperature evolution of the superconducting gap distribution in Pb$_{0.64}$Bi$_{0.36}$ alloy, calculated within the anisotropic SCDFT approach.}\label{fig:aniso2}
\end{figure}

The situation is more complex for the case of the Pb$_{0.64}$Bi$_{0.36}$ alloy, which is directly related to the more complex fermiology, as shown in Fig.~\ref{fig:aniso-pbbi}.
The stronger electron-phonon interaction results in the $\lambda_{n\mathbf{k}}$ spanning the range from 1.45 to 2.51, with the average values per each FS sheet of 1.92, 2.03 and 2.22 [Fig.~\ref{fig:aniso-pbbi}(a-d)].
The global average $\lambda = 2.08$ agrees well with 2.05 from the isotropic Eliashberg function. 
In contrast to Pb, the histogram of $\lambda_{n\mathbf{k}}$ shows the overlap of the values between the FS sheets. 
The screened Coulomb pseudopotential $\mu_{n\mathbf{k}}$
shown in Fig.~\ref{fig:aniso-pbbi}(e-h) has a rather flat distribution inside the three FS sheets, with increasing averages from 0.302 to 0.311. This gives a triangular-like distribution for the global value with the average $\mu = 0.30$ (Table~\ref{tab:sctk}), 7\% higher than in Pb.
The superconducting gap $\Delta_{n\mathbf{k}}$ [Fig.~\ref{fig:aniso-pbbi}(i-l)] spans the range from 1.68 meV to 2.05 meV (which translates to $\frac{2\Delta}{k_BT_c}$ in the range of 4.06--4.95) and shows more overlapped behavior than in Pb. 
The largest gap values are found on the third sheet of the Fermi surface,  with the average $\langle\lambda_3\rangle = 2.22$ and $\langle{\Delta}_3\rangle = 1.93$ meV. 
The other two pieces of the Fermi surface are more isotropic, with regions of weaker EPC and smaller gaps (see Table~\ref{tab:sctk}). 
These are the areas where the hybridization between $p_{1/2}$ and $s$ is the strongest (see Fig. \ref{fig:elsoc}(e-g)). 
The band inversion, which mixes the s/p band character and increases the p-orbital contribution to the largest (second) FS sheet, smears the superconducting gap distribution over the FS sheets, and as a consequence we do not observe separated gap histograms as for Pb. 
Instead, we have a strongly anisotropic gap distribution with three local maxima, as shown in Fig.~\ref{fig:aniso-pbbi}(l). 
This gap structure with a significant spread of the values and anisotropy is responsible for the deviation of the temperature dependence of the thermodynamic properties from the isotropic Eliashberg picture. 
As the temperature increases, the gap structure changes, as shown in Fig.~\ref{fig:aniso2}. The spread in gap values becomes narrower, which explains why deviations from the single-band s-wave Eliashberg formalism were mostly observed in the lower temperature range.

\subsubsection{The retarded screened Coulomb interaction}

Before moving on to the effect of disorder on $T_c$, we discuss the Coulomb interaction parameters $\mu$ and $\mu^*$ in more detail. The screened Coulomb interaction parameter $\mu$, as mentioned above, is computed from the electron-electron interaction kernel~\cite{sctk}.
In the superconducting state, the effect of Coulomb repulsive interactions is effectively weakened because of the retardation effects between phonons and electrons, and this effect is taken naturally into account in the SCDFT~\cite{scdft-metals}.
In the standard Eliashberg approach or in the McMillan-Allen-Dynes formulas,
retardation is approximately described using the $\mu^*$ parameter~\cite{Bogoljubov1958,Morel1962,mcmillan,allen-dynes,Carbotte1990,chang_nb,gunnarsson2013}
\begin{equation}\label{eq:mu*}
    \mu^* = \frac{\mu}{1+\mu\ \ln(\omega_{\rm el}/\omega_{\rm ph})}.
\end{equation}
The $\omega_{\rm el}$ and $\omega_{\rm ph}$ are the characteristic electronic and phononic frequencies (energies), of the order of Fermi energy or bandwidth for the former and the highest phonon frequency $\omega_{\rm max}$ or the Debye $\omega_D$ for the latter.
Allen and Dynes~\cite{allen-dynes} used the plasmon frequency $\omega_p = \omega_{\rm el}$ for electrons and the maximum frequency $\omega_{\rm max}$ as $\omega_{\rm ph}$ for phonons.
Note here that because $\mu^*$ depends on the maximum phonon frequency, once it is used to solve the Eliashberg equations it depends on the phonon cutoff frequency $\omega_c$. In our case  $\omega_c = 8\times\omega_{\rm max}$, this is responsible for the fact that different $\mu^*$ are used in both approaches (i.e. numerical solution of isotropic Eliashberg equations and Allen-Dynes formula) to obtain the same $T_c$, and when they are to be compared, the former $\mu^*$ value has to be re-scaled based on the $\omega_c/\omega_{\rm max}$ ratio~\cite{allen-dynes,kuderowicz2021}:
$\frac{1}{\widetilde{\mu^*}} = \frac{1}{\mu^*} + \ln\left(\frac{\omega_c}{\omega_{\rm max}}\right)$, where $\widetilde{\mu^*}$ can be compared to $\mu^*$ used in the Allen-Dynes formula or computed with Eq.(\ref{eq:mu*}).

As we discussed above, to obtain the experimental critical temperatures in the isotropic Eliashberg approach, we used $\mu^*=0.117$ (Pb) and $0.134$ (Pb-Bi), which for $\omega_c = 8\omega_{\rm max}$ give the re-scaled values of $\widetilde{\mu^*} = 0.094$ and 0.105, respectively. So both are close to the common value of 0.10, but note the 10\% higher value for Pb-Bi is required to reach $T_{c, \rm expt}$. 

These are to be confronted with the first-principles values of $\mu$. The calculated values of the screened Coulomb interaction parameters $\mu_{n\mathbf{k}}$ on the Fermi surface show an approximately 10\% spread for both Pb and Pb-Bi, as presented in Fig.~\ref{fig:aniso-pb}(d-f) and Fig.~\ref{fig:aniso-pbbi}(e-h). 
The average values are $\mu=0.27$ for Pb and $\mu=0.30$ for Pb-Bi.
To obtain the retarded values according to Eq.(\ref{eq:mu*}) we computed the dielectric function tensor $\epsilon_{\alpha\beta}(\omega)$ using QE and from this the plasmon frequency was calculated as $\omega_p=\frac{1}{3}\sum_{\alpha=1}^{3}\frac{2}{\pi}\int \omega \Im(\epsilon_{\alpha\alpha}(\omega)) d\omega$.
For Pb, we obtain $\omega_p=9.9$ eV, and for Pb-Bi $\omega_p=10.0$ eV.
In combination with the maximum phonon frequencies, that gives $\mu^*=0.093$ for Pb and $\mu^*=0.096$ for the Pb-Bi alloy. 
For Pb, all approaches converge to the same $T_c$ and $\mu^*$ values, well corresponding with the experimental observations. 
On the other hand, for the Pb-Bi alloy, 
the calculated Coulomb repulsion parameter $\mu^*$ is only 3\% larger than in Pb.
The overestimated $T_c = 9.6$~K obtained in SCDFT, as well as the larger $\widetilde{\mu^*} = 0.105$ required in the Eliashberg approach to obtain $T_c = T_{c, \rm expt}$ strongly suggest that in real Pb-Bi samples the critical temperature is reduced by an additional mechanism beyond the Coulomb interactions. 
We propose this mechanism to be a disordered-induced electron scattering, which was not included in the studies mentioned above and is discussed in the next paragraph.

\begin{table*}[t]
\caption{The essential parameters for superconductivity of Pb and Pb$_{0.64}$Bi$_{0.36}$ obtained with SCDFT, together with experimental data, taken from this work and \cite{allen-dynes,pb-gap,pbbi-history5}. $\lambda$, $\mu$ and $\Delta(0)$ are the average electron-phonon coupling constant, screened Coulomb repulsion parameter, and the superconducting gap at $T = 0$~K, respectively, while $\langle \lambda_i\rangle$, $\langle \mu_i\rangle$ and  $\langle \Delta_i(0)\rangle$ are the average values on the Fermi surface sheet $i$. The superconducting gap is given in meV, and the critical temperature is given in K. {The data obtained from tunneling measurements for an alloy with a very similar concentration (Pb$_{0.65}$Bi$_{0.35}$) are included for comparison.}}
    \centering
    \begin{ruledtabular}
    \begin{tabular}{l c c c c c  c c c c c  c c c c }
 & $\lambda$ &$\langle\lambda_1\rangle$&$\langle\lambda_2\rangle$&$\langle\lambda_3\rangle$   &     $\mu$ &$\langle\mu_1\rangle$&$\langle\mu_2\rangle$&$\langle\mu_3\rangle$&  $ \mu^* $&     $ \Delta(0) $ &$\langle\Delta_1(0)\rangle$&$\langle\Delta_2(0)\rangle$&$\langle\Delta_3(0)\rangle$&        $T_c$\\
    \hline
     Pb - SCDFT    & 1.59&1.40&1.70&---& 0.271  &0.263&0.274& ---&0.093&    1.39& 1.30 & 1.42 &---&7.3\\ 
     Pb - expt.\footnote{\citet{pbbi-history5}\label{fr}}   &1.55\footnote{\citet{pb-gap}}-1.60\footnote{\citet{allen-dynes}\label{fr2}}&&&&&&&& 0.105\footref{fr2}& 1.36 &1.16--1.28 &1.37--1.43&---& 7.2\\
Pb$_{0.64}$Bi$_{0.36}$ - SCDFT& 2.08&1.92 & 2.03 & 2.22 &  0.307  &0.302&0.307&0.311 &0.096 & 1.88&1.84 &1.88&1.93 & 9.6 \\
Pb$_{0.64}$Bi$_{0.36}$ - expt. & 2.09\footnote{calculated from Eqs. (\ref{eq:tc-modified}-\ref{eq:tc-f2}), using experimentally determined $T_c$ with $\mu^*=0.1$; the value obtained from Eq. \ref{eq:gamma-lambda} using heat capacity data  is equal to 1.83 or 1.96, depending on the electronic band structure.}& &&  &   &&&  &  & && & &8.6  \\
Pb$_{0.65}$Bi$_{0.35}$ - expt. \footref{fr2} & 2.13& &&  &   &&&  & 0.093&1.84&& & &8.95  \\
    \end{tabular}
    \end{ruledtabular}
    \label{tab:sctk}
\end{table*}

\subsubsection{Effect of disorder on $T_c$}
The final question we would like to address in this work is whether we observe the suppressing effect of disorder on the critical temperature in Pb-Bi alloy.
According to Anderson's theorem ~\cite{anderson_59}, conventional superconductivity remains unaffected by weak disorder caused by nonmagnetic impurities.
However, in strongly disordered cases, suppression of superconductivity is expected~\cite{anderson_degradation,weakly_localized_regime}, as observed in A-15 superconductors~\cite{anderson_degradation},  highly disordered metals~\cite{destruction_granular} or thin films~\cite{destruction_films}.
In the simplest approach, this effect can be captured as an increase in the $\mu^*$ parameter value required to reproduce experimental $T_c$~\cite{anderson_degradation}, which now will contain the depairing Coulomb interactions and the effect of scattering.
Very strong electron scattering, on the border of the Mott-Ioffe-Regel limit, was recently suggested to be responsible for the decrease in $T_c$ in (ScZrNb)$_{1-x}$RhPd$_x$ series of superconducting alloys~\cite{gutowska2023}. 
There, the experimentally determined critical temperature was found to decrease significantly with increasing $x$, contrary to predictions based on calculations of the electronic contribution to $\lambda$~\cite{gutowska2023}.
Electron scattering, induced by disorder, affects also other thermodynamic properties of the material, like the specific heat jump at $T_c$, $\Delta C/\gamma T_c$ which can be reduced in weakly-coupled materials below the BCS value of 1.43~\cite{heat-prb}.
However, the strongest renormalization due to scattering concerns the upper magnetic critical field~\cite{Carbotte1990}, which for the given $T_c$ increases with increasing scattering rate, as we observed in Fig.~\ref{fig:Hc2}.

\begin{figure}[b] 
\includegraphics[width=0.50\textwidth]{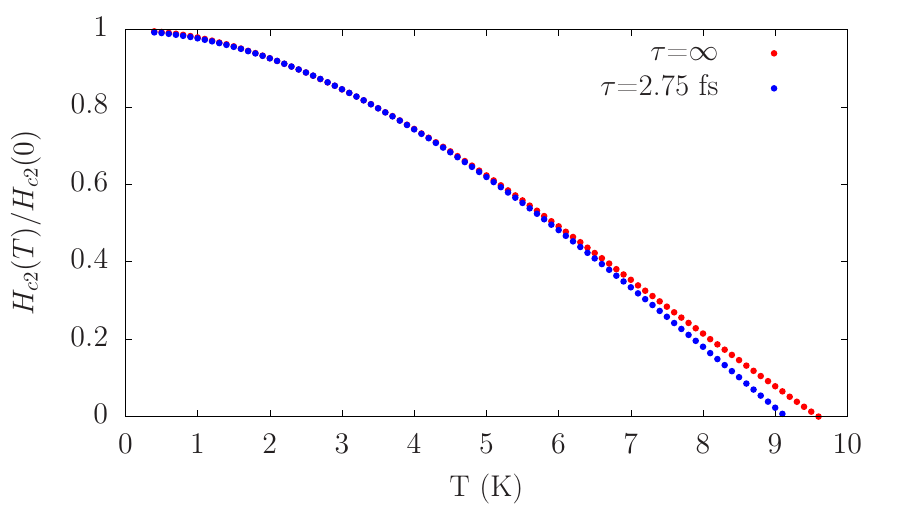}
\caption{Effect of disorder-induced electron scattering on the reduced $H_{c2}$ and critical temperature.}\label{fig:hc2-tau}
\end{figure}

In our case of Pb$_{0.64}$Bi$_{0.36}$ the mean-free path $\ell$ estimated for $\tau = 2.75$ fs is $\ell = 28$~\AA. Thus, despite $\tau$ being short, due to the high Fermi velocity $\ell$ is much above the interatomic distance. 
The scattering time below which one can expect the suppressing effect disorder on superconductivity is usually given by the criterion~\cite{weakly_localized_regime} $\hbar/\tau \gg \hbar\omega_D$, where the Debye frequency sets the characteristic phonon energy scale (i.e. the scattering time is much shorter than the period of atomic vibrations). In our case, this condition is satisfied by far, as $\hbar/\tau = 240$~meV $\gg  \hbar\omega_D = 9$~meV. This conclusion is not changed if the maximum phonon frequency is used instead of $\omega_D$.

To analyze the effect of scattering on $T_c$ quantitatively, we return to the isotropic Eliashberg formalism and recalculated the $H_{c2}(T)$ curves to capture the effect of scattering on $T_c$.
First, we took the parameter-free critical temperature obtained in the SCDFT calculations, where electron scattering is not taken into account, $T_c = 9.6$~ K, and calculated the $H_{c2}(T)$ curve to match the same $T_{c0} = 9.6$~ K without scattering (the scattering time $\tau = \infty$, clean limit). 
That required taking $\mu_0^* = 0.075$. Next, using the same $\mu_0^*$, the $H_{c2}(T)$ curve was recalculated including the scattering effects with $\tau = 2.75$ fs. 
That resulted in a reduction of the critical temperature to $T_c = 9.1$~K, as shown in Fig.~\ref{fig:hc2-tau}.
Although the isotropic formalism does not reproduce the precise reduction to experimental $T_{c} = 8.6$~ K, it confirms that for this level of the scattering time $\tau$ the reduction in the critical temperature is observed.

\section{Summary and conclusions}
In summary, the electronic structure, phonons, electron-phonon interaction, and superconductivity were studied for the hexagonal Pb$_{0.64}$Bi$_{0.36}$ alloy in relation to cubic Pb. 
The cubic-to-hexagonal phase transition was explained by the energetically unfavorable effect of an increase in the occupation of antibonding states in the cubic phase. 
The electronic structure of the alloy was found to be quite strongly smeared by disorder, with a scattering time of $\tau \sim 3$ fs. Nevertheless, the virtual crystal approximation within the mixed pseudopotential scheme was found to be suitable to describe the electron-phonon interactions in the material. The record strong electron-phonon coupling in the Pb-Bi alloy is the result of a combination of several favorable factors. The major increase in $\lambda$ is driven by the cubic-to-hexagonal transition, which creates the layered structure and lowers the phonon frequencies of the out-of-phase vibrations of hexagonal atomic layers, stacked along the $c$ axis. 
This promotes a stronger electron-phonon coupling. The electron doping effect additionally increases the electronic contribution to $\lambda$, and both effects result in an increase of $\lambda$ from 1.47 in Pb to 2.08 in Pb$_{0.64}$Bi$_{0.36}$.
The strong spin-orbit coupling also cooperates to boost the $\lambda$, as due to relativistic $p$-orbital contraction SOC lowers the phonon frequencies. 
As a result, strongly coupled superconductivity emerges, with 
$\frac{2\Delta}{k_BT_c}$ being in the range from 4 to 5
and $\frac{\Delta C}{\gamma T_c} = 2.9$.
Due to a complex Fermi surface with three large sheets of different levels of $sp$-hybridization, followed by anisotropy of the electron-phonon interaction, strongly anisotropic three-band superconductivity is realized, with three overlapping maxima in the superconducting gap function in the reciprocal space.
This anisotropic nature of the superconducting state is confirmed both by theory and experiment, where the magnetic critical fields and specific heat in the superconducting state were found to deviate from the single-band isotropic Eliashberg theory.
Furthermore, the strong electron scattering effect on $T_c$ was analyzed and found to lower the magnitude of the critical temperature in Pb$_{0.64}$Bi$_{0.36}$.

\section*{Acknowledgments}
The work at AGH University was supported by the National Science Centre (Poland), project number 2017/26/E/ST3/00119 and by the Polish high-performance computing infrastructure PLGrid (HPC Center: ACK Cyfronet AGH) by providing computer facilities and support within computational grants no. PLG/2023/016451 and PLG/2024/017305.
K.G. acknowledges support from the U.S. Department of Energy, Office of Science, Basic Energy Sciences, Materials Sciences and Engineering Division.
The work in Gdansk was supported by the National Science Centre (Poland), Grant no. 2022/45/B/ST5/03916.

\bibliography{references}

\begin{thebibliography}{90}
\bibitem [{pse({\natexlab{a}})}]{pseudo}%
  \BibitemOpen
  \href@noop {} {} ({\natexlab{a}}),\ \bibinfo {note} {used pseudopotentials: Pb.rel-pbe-dn-rrkjus\_psl.1.0.0.UPF and Bi.rel-pbe-dn-rrkjus\_psl.1.0.0.UPF, which have been generated with the help of QE, using input files from PSlibrary package \cite{pps}}\BibitemShut {NoStop}%
\bibitem [{pse({\natexlab{b}})}]{pseudo2}%
  \BibitemOpen
  \href@noop {} {} ({\natexlab{b}}),\ \bibinfo {note} {in this case we used pseudopotentials generated using input files Bi.rel-pz-n-nc.UPF and Bi.rel-pz-n-nc.UPF.in from PSlibrary package \cite{pps}}\BibitemShut {NoStop}%
\end{thebibliography}

\vspace*{24pt}

\newpage
\onecolumngrid
\section*{Supplemental Material}

\renewcommand{\thefigure}{{S\arabic{figure}}}
\setcounter{figure} 0

\vspace*{24pt}
\noindent
Supplemental Material contains:
\begin{enumerate}
\item Band structure of Pb-Bi alloy with marked location of k-points used in Fig.~\ref{fig:bsf2} in the main text.
\item Definition of the parameter $I$ and definition of the electron-phonon parameter $\lambda$ with the use of $I$.
\item Analysis of the force constant values of Pb and Pb-Bi and its correspondence to atomic bonding.
\item Additional description of the isotropic Eliashberg equations.
\item Phonon and electron-phonon properties of Pb calculated in a scalar-relativistic way.
\item Remarks on the Peierls distortion in Bi.
\end{enumerate}

\section{k-points used in BSF analysis}
\begin{figure}[h]
\includegraphics[width=.5\textwidth]{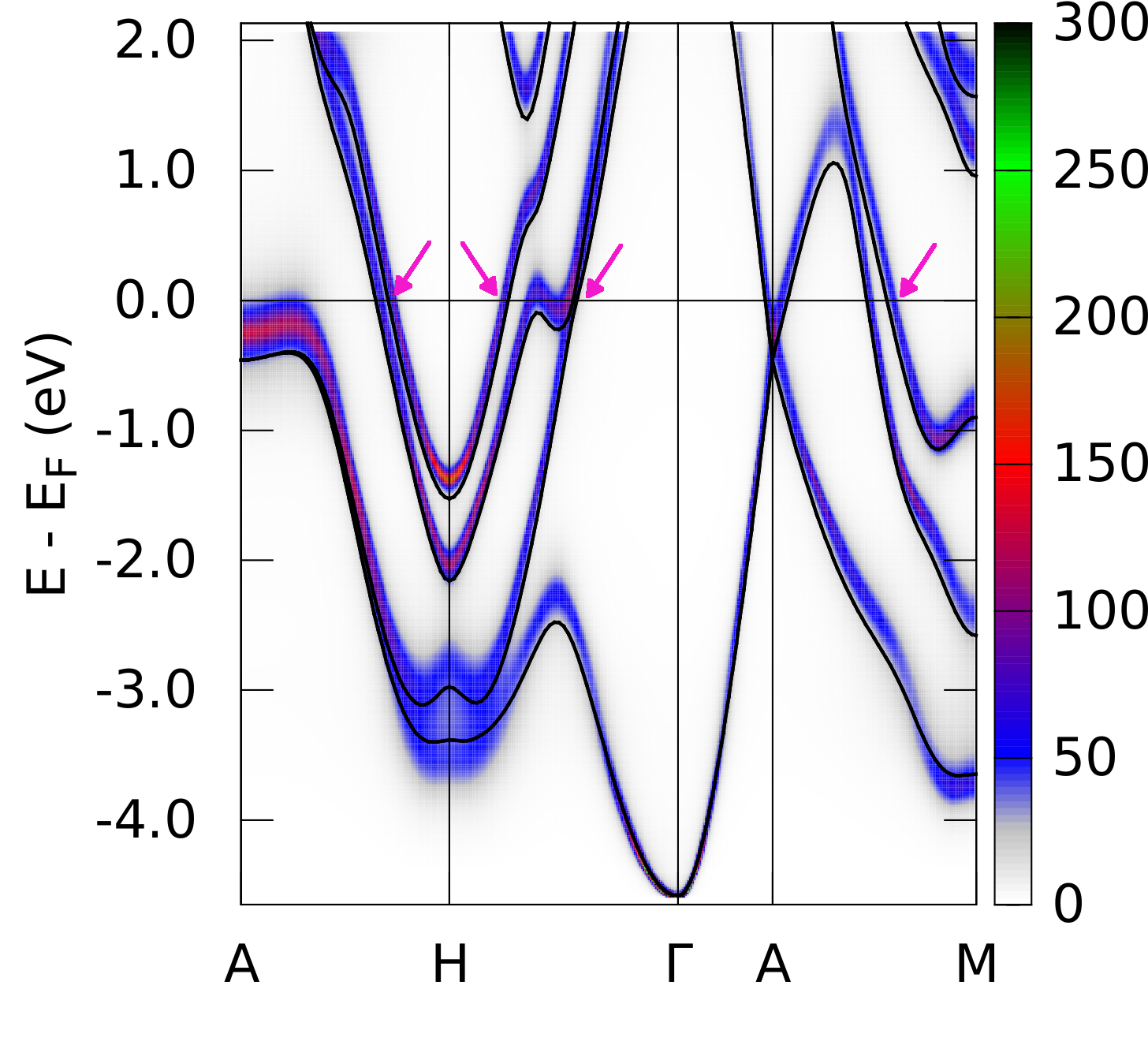}
\caption{Location of $\mathbf{k}$-points used in Fig. 12 in the main manuscript.}\label{fig:bsf-location}
\end{figure}

Arrows in Fig.~\ref{fig:bsf-location} show the location of $\mathbf{k}$-points used in Fig. 12 in the main manuscript to plot the energy-dependent BSF: one point in A-H direction, two points in H-$\Gamma$ and one point in A-M. 

\section{parameter I}
We define the integral $I$ as:
\begin{equation}
I=\int_0^{\omega_{\rm max}} \omega\cdot\alpha^2F(\omega){\rm d}\omega.
\label{eq:I}
\end{equation}
which is a frequency-independent measure of  electronic contribution to $\lambda$~\cite{gutowska2021,kuderowicz2022}.
This quantity does not depend directly on phonon frequency, as multiplication over $\omega$ cancels the $\alpha^2F(\omega) \propto 1/\omega$ dependence, and $I$ is proportional to the sum of phonon linewidths over the Brillouin zone, directly dependent on the electronic wave function matrix elements: 
\begin{equation}
\begin{split}
I&=\frac{1}{2\pi\hbar N(E_F)} \int_0^{\omega_{\rm max}} d\omega
                    \sum_{{\bf q}\nu}
                    \delta(\omega-\omega_{{\bf q}\nu}) \gamma_{\textbf{q}\nu}=\\
& \frac{1}{2\pi\hbar N(E_F)} \int_0^{\omega_{\rm max}} d\omega
                    \sum_{{\bf q}\nu}
                    \delta(\omega-\omega_{{\bf q}\nu})
                     \sum_s \frac{1}{2M_s} 
\int \frac{d^3k}{\Omega_{BZ}}|\langle\psi_{i,{\bf k+q}}| {dV_{\rm SCF}\over d {\hat u}_{\nu s} }\cdot
                   \hat \epsilon_{\nu s}|\psi_{j,{\bf k}}\rangle|^2 
\delta(E_{{\bf k},i} - E_F)  \delta(E_{{\bf k+q},j} - E_F),\\
\end{split}
\end{equation} 
thus
\begin{equation}
I=\frac{1}{2\pi\hbar N(E_F)} 
\sum_{{\bf q}\nu}
\gamma_{\textbf{q}\nu}.
\end{equation}     

With the help of $I$ and the "average square" phonon frequency, defined as 
\begin{equation}
    \hat\omega^2 = \int_0^{\omega_{\rm max}} \omega \alpha^2F d\omega
    \left/ \int_0^{\omega_{\rm max}} \alpha^2F \frac{d\omega}{\omega}\right.
\end{equation}
EPC constant is expressed as 
\begin{equation}
\lambda = \frac{2I}{\hat\omega^2}
\end{equation}
i.e. it is the ratio of mainly electronic and phononic contributions to $\lambda$.
For a system where $\alpha^2F(\omega) \simeq const \times F(\omega)$, the parameter $\hat\omega^2$  primarily depends on the shape of phonon DOS.
In a monoatomic system the McMillan-Hopfield parameter~\cite{mcmillan,hopfield} is defined as $\eta = 2MI$ ($M$ is the atomic mass) and then the electron-phonon coupling constant is calculated using the well-known formula:
\begin{equation}
\lambda = \frac{\eta}{M\hat\omega^2}.      
\end{equation}

\section{Force constants analysis}

\begin{figure}[H]
    \centering
\includegraphics[width=0.8\linewidth]{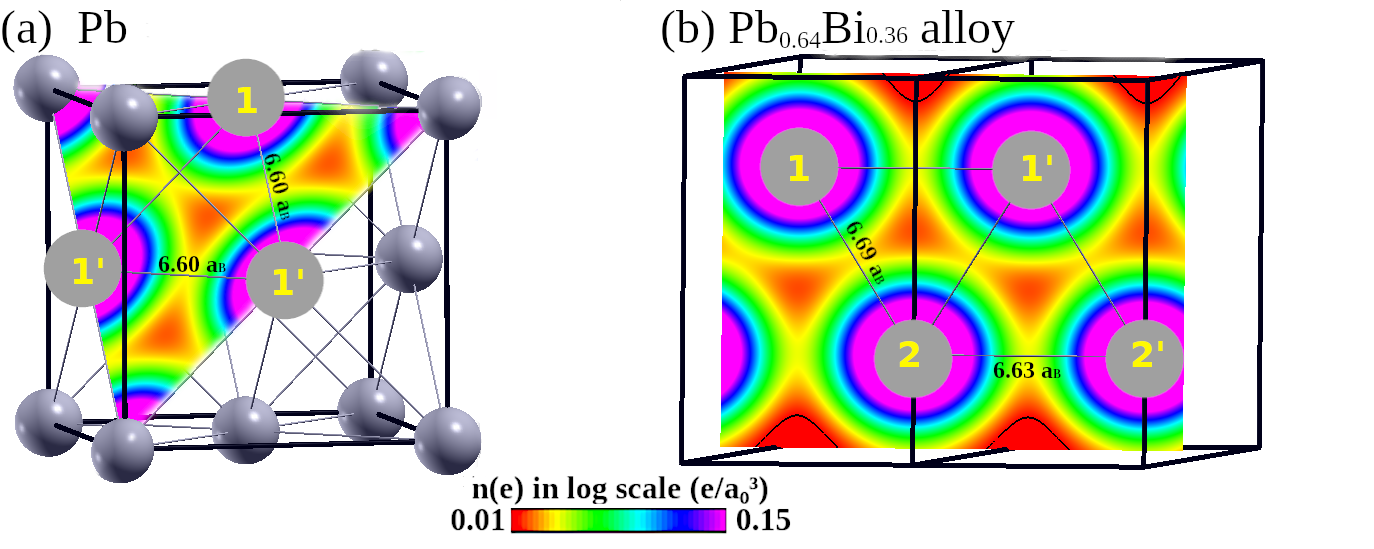}
    \caption{The conventional unit cell of Pb and 2x1x1 supercell of Pb-Bi alloy. Atom 1' is equivalent to atom 1, while atom 2 is in the next layer.} 
    \label{fig:suplat}
\end{figure}

To understand why the frequencies are lowered due to the $fcc$-$hcp$ transition, we have to look at the force constants in these two variants of Pb. 
It is worth considering the effective force constants $g_m$ acting on a given atom in the $m$ direction, when it is displaced in that direction (this force constant describes the total restoring force acting on atom after its displacement).
The values are shown in Table \ref{tab:fc}. 
When the structure changes from $fcc$ to $hcp$, the atomic layers are formed and the atomic distances in the plane (between atoms 1 and 1' in Fig.~\ref{fig:suplat}) and between the planes (between atoms 1 and 2 in Fig.~\ref{fig:suplat}) increase.
Atom-atom bonds become weaker, leading to a smaller force constant, especially in the $z$ direction, as the $hcp$ is elongated in this direction. 
The main effects for the restoring force are the following:
\begin{itemize}
\item $fcc$-$hcp$ distortion in Pb lowers the force constant $g$;
\item alloying with Bi (electron doping) increases the charge density and slightly increases $g$ due to long-range effects, as the nearest-neighbors force constants slightly decrease;
\item Spin-orbit coupling in Pb-Bi significantly decreases $g$ due to the $p$-orbital contraction, which lowers the nearest-neighbor force constants.
\end{itemize}
As a consequence, the phonon frequencies are lowered.

\begin{table}[htb!] 
\caption{"Stiffness" $g$ - the effective force constants expressed in mRy/$a_B$. Stiffness $g_{m}$ is a constant of a restoring force acting on atom in direction $m$ when it is displaced in the same direction. An additional value $g=\sqrt{g_x^2+g_y^2+g_z^2}$ is calculated.}
\label{tab:fc}
\begin{center}
\begin{ruledtabular}
\begin{tabular}{ c c c c c}
		&$g_{x}$ &		$g_{y}$&			$g_{z}$ & $g$	\\
fcc Pb	&	37.3	&	37.3	&	37.3	&	64.6	\\
hcp Pb	&	32.3	&	32.3	&	28.5	&	53.9	\\
Pb$_{0.64}$Bi$_{0.36}$ 	&	31.7	&	31.7	&	31.2	&	54.5		\\
Pb$_{0.64}$Bi$_{0.36}$ w/o SOC	&	36.6	&	36.6	&	36.5	&	63.3	\\
\end{tabular}
\end{ruledtabular}
\end{center}
\end{table}

\section{Eliashberg formalism}
Eliashberg equations, defined in the imaginary axis, are given by~\cite{Eliashberg1960,kuderowicz2021}
\begin{eqnarray}
Z(i\omega _n)&=&1+\frac{\pi k_B T}{\omega _n} \sum _{n'} \frac{\omega _{n'}}{R(i\omega _{n'})} K(i(\omega_{n}-\omega_{n'}) \nonumber \\
 Z(i\omega _n) \Delta(i\omega _n) &=& \pi k_B T \sum _{n'} \frac{\Delta(i\omega _{n'})}{R(i\omega _{n'})} \times 
 [ K(i(\omega_n-\omega_{n'}))-\mu^* \theta ( \omega _c - |\omega _{n'}| )],
 \label{eq:eliash}
\end{eqnarray}
where $Z(i\omega_n)$ is the mass renormalization function, $\Delta(i\omega _n)$ is the superconducting order parameter, $i\omega_n=i(2n+1)\pi k_B T$ 
are fermionic Matsubara frequencies where $n\in \mathbb{Z}$, $\theta(\omega)$ is the Heviside function, $k_B$ is the Boltzmann constant, $T$ is  
temperature and 
\begin{equation}
R(i\omega _n) = \sqrt{\omega _n ^2 + \Delta^2(i\omega _n)}. 
\end{equation}
The kernel of the electron-phonon interaction takes the form of
\begin{equation}
 K(i(\omega_n-\omega_{n'})=\int _0^{\infty} d \omega \frac{2 \omega \alpha ^2 F(\omega)}{(\omega _n - \omega _{n'})^2+\omega ^2},
\end{equation}
where $\alpha ^2 F(\omega)$ is the isotropic Eliashberg spectral function.
Equations (\ref{eq:eliash}) are solved in a self-consistent way using the following parameters: number of Matsubara frequencies $M=8000$ and cutoff energy $\omega_c=8\omega_{max}$, where $\omega_{max}$ is the upper limit of the phonon frequency.
The pseudopotential parameter $\mu^*$ describes the depairing Coulomb electron-electron interactions. 

Once the equations are solved, 
the difference in the electronic specific heat determined in the superconducting and normal state $\Delta C_e=C_e^S-C_e^N$ can be expressed as
\begin{equation}
 \frac{\Delta C_e(T)}{k_B N(E_F)} = -\frac{1}{\beta}\frac{d^2 \Delta F /N(E_F)}{d(k_BT)^2},
\label{eq:Cv}
\end{equation}
with the specific heat in the normal state given by
\begin{equation}
\frac{C_e^N(T)}{k_B N(E_F)}=\frac{\pi^2}{3} k_B T (1+\lambda),
\end{equation}
where $\lambda$ is the electron-phonon coupling constant and $N(E_F)$ corresponds to the density of states at the Fermi level.
In Eq.(\ref{eq:Cv}), $\Delta F$ is the free energy difference between the superconducting and normal state 
which takes the form 
\begin{eqnarray}
 \frac{\Delta F}{N(E_F)}=-\pi k_BT \sum _{n} \left ( \sqrt{\omega_n^2+\Delta _n^2} - |\omega _n|\right ) \times
  \left ( Z^S(i\omega _n) - Z^N(i\omega _n) \frac{|\omega _n|}{\sqrt{\omega _n^2+\Delta_n^2}} \right ),
\end{eqnarray}
where $Z^S$ and $Z^N$ denote the mass renormalization factors for the superconducting (S) and normal (N) states, respectively.

The upper magnetic critical field is calculated on the basis of equations ~\cite{Carbotte1990,kuderowicz2021}:
\begin{equation}
    \tilde{\Delta}(i\omega _n)=\pi k_BT \sum _{n'} \frac{\left [ \lambda(n-n')-\mu^* \theta(\omega_c - |\omega _{n'}|) \right ] \tilde{\Delta}(i\omega _{n'}) }{\chi ^{-1}[\tilde{\omega}(i\omega_{n'})]-\pi t^+},
\label{eq:delta_hc2}
\end{equation}
where
\begin{equation}
\tilde{\omega} (i\omega_{n})=\omega _n + \pi k_BT \sum _{n'}\lambda(n-n') \text{sgn}(\omega_{n'})+\pi t^+ \text{sgn}(\omega_{n'}),
\label{eq:omega_hc2}
\end{equation}
with $t^+=\frac{1}{2\pi\tau}$, where $\tau$ is the electronic scattering time. 
In Eq.~(\ref{eq:delta_hc2}), the function $\chi(\tilde{\omega})$ is given by the formula
\begin{equation}
    \chi (\tilde{\omega}(i\omega _n))=\frac{2}{\sqrt{\alpha}} \int _0 ^\infty dq \: e^{-q^2} \tan^{-1} \left ( \frac{\sqrt{\alpha}q}{|\tilde{\omega}(i\omega _n)|} \right ),
\end{equation}
with 
\begin{equation}
\alpha(T)=\frac{1}{2}|e|H_{c2}(T)v_F^2,   
\end{equation}
where $e$ is the elementary charge and $v_F$ is the Fermi velocity.

\section{Phonons and $\lambda$ of Pb calculated in a scalar-relativistic way}

\begin{figure} \centering \includegraphics[width=0.5\linewidth]{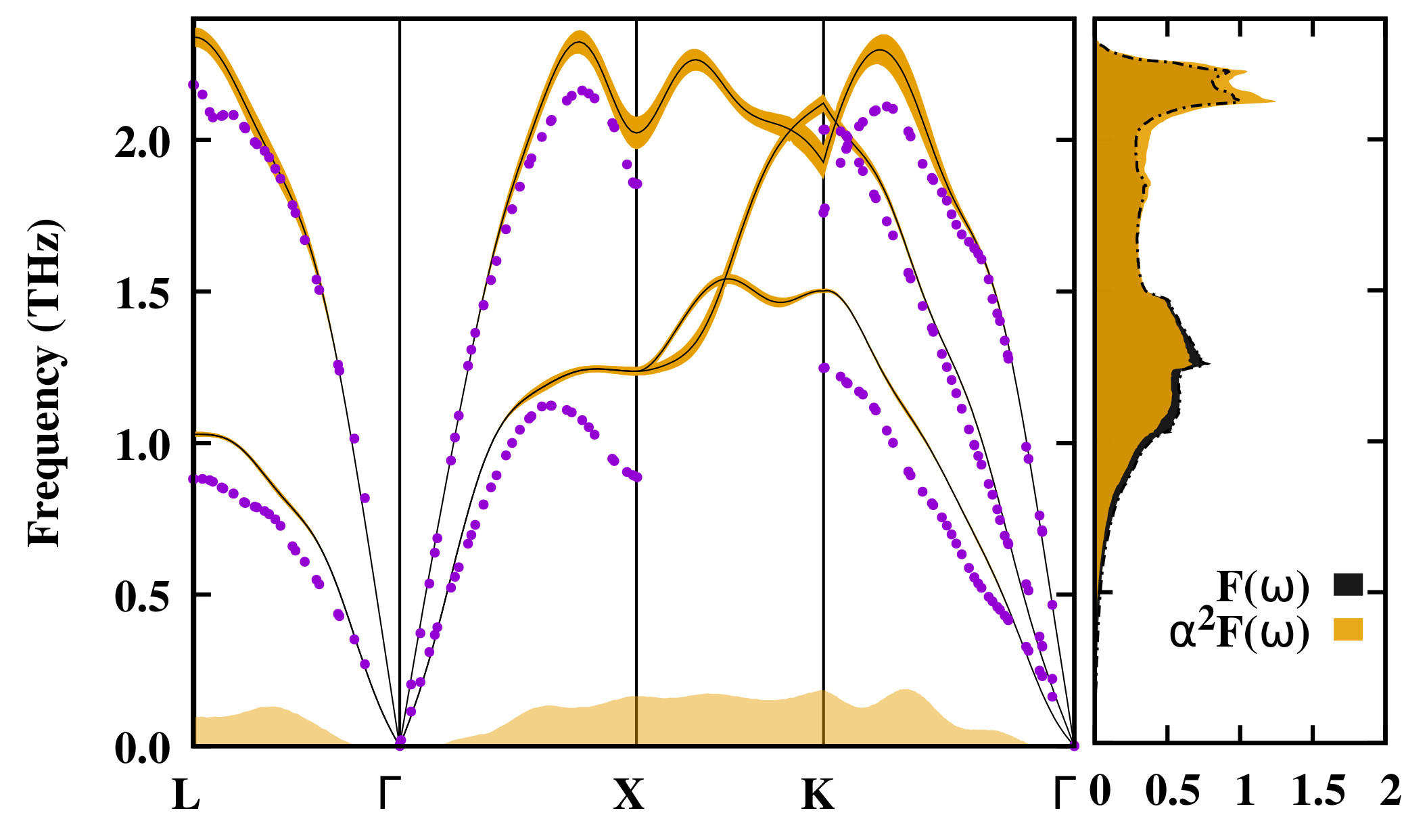} \caption{Phonon and electron-phonon properties of Pb in the scalar-relativistic case: (a) dispersion relation, with phonon linewidth indicated by yellow shading and experimental data marked with dots;  (b) phonon DOS together with the Eliashberg function.} 
\label{fig:pb-ph} 
\end{figure}

In the main text, we describe the phonon and electron-phonon properties of Pb calculated in a fully relativistic manner, which includes all relativistic effects. In contrast to that, here we show the properties of Pb calculated is scalar-relativistic way, which means, that the effect of spin-orbit coupling (SOC) is neglected. As seen in Fig. 7 of the main text, where the electronic structure of Pb is shown for both fully and scalar-relativistic cases, the effect of SOC leads to noticeable changes. Specifically, SOC contracts the $p$-orbitals, as illustrated in Fig. 9 of the main text, which shows the difference in electronic charge density calculated with and without SOC. In such a case, we expect that atomic bonds are stiffer in the scalar-relativistic case than in the fully relativistic case, leading to higher (overestimated) phonon frequencies in the scalar-relativistic case. This expectation is confirmed in Fig. \ref{fig:pb-ph}, where the phonons calculated without including SOC in the electronic structure are shown. Indeed, the phonon frequencies are overestimated compared to the experimental data, which contrasts with the fully relativistic case shown in the main text. 

As a consequence, the predicted electron-phonon coupling constant is underestimated (equal to 1.0 instead of 1.5), leading to an underestimation of the critical temperature (5 K instead of the experimentally measured 7 K; see Table IV in the main text).
This underlines the significant role of spin-orbit coupling  in the case of Pb.

\section{Peierls distortion in Bi}

\begin{figure*}[h]
\includegraphics[width=\textwidth]{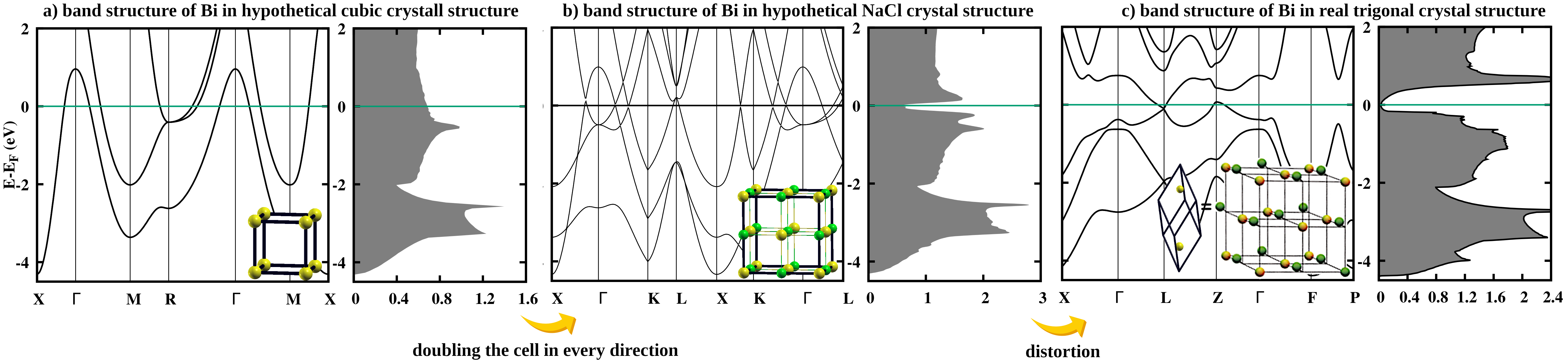}
\caption{The crystal unit cell and electronic structure in terms of bands and DOS of Bi in (a) non-existing cubic crystal structure, (b) non-existing NaCl $fcc$ structure and (c) its real trigonal structure. The evolution of crystal structure is shown in subpanels. In (a) there is one-atomic cubic cell. By doubling this cell in every direction and distinguishing half of atoms, the NaCl cell arises, shown in (b). When one half of atoms is distorted (as shown in (c)), the trigonal cell of Bi arises, which is shown in (c). The lattice constants of (a) and (b) are chosen such that the nearest neighboor's distances are equal to average of distances of nearest and second nearest neigboors in (c).}\label{fig:bi}
\end{figure*}

The crystal structure of a solid Bi is trigonal (rhombohedral, $R$-$3m$)
which may be alternatively described in a hexagonal unit cell, which is a tripled rhombohedral primitive cell.
It is formed from the $fcc$ NaCl-type of crystal structure, distorted along the diagonal due to the Peierls distortion (see Fig. \ref{fig:bi}). 
The distortion creates the diatomic primitive cell and appears as Bi has half of the $p$-shell occupied (Bi: $6s^26p^3$). In a diatomic unit cell the Bi-Bi bonds form bonding and antibonding orbitals, with the occupied bonding lowered in energy, while the antibonding orbitals are pushed to the higher energies. 
The energy gain caused by lowering the electronic states is larger than the energy needed to make a distortion, making this distorted structure more stable \cite{peierls}. And due to the even number of six $p$ electrons, the bonding orbitals may be (almost) fully occupied, leaving the antibonding (almost) empty. Almost, as solid Bi is a semimetal with a small overlap of the valence and conduction bands.

In the simplified one-dimensional band structure picture, by doubling the unit cell, the Brillouin zone is reduced by a factor of two, and at the boundary of the new Brillouin zone, where the original Fermi wavee vector was located, the gap opens. This pushes the occupied electronic states to lower energies, making this material insulating or semi-metallic \cite{peierls2}. 

The case of Bi is shown in Fig. \ref{fig:bi}, where the band structure of Bi in simple cubic cell, NaCl structure and trigonal structure are shown together with the pictures of those structures. If Bi were cubic, it would have a metallic character. When the structure is changed to NaCl-type (with two nonequivalent Bi atoms), the Brillouin zone is reduced by a factor of two in every direction, causing many Dirac-like cones at the Fermi level and the valley in DOS appears. When the structure is further distorted to a trigonal structure, most of the Dirac cones are gapped, and Bi becomes semi-metallic.

\end{document}